\documentclass[fleqn,usenatbib]{mnras}
\usepackage{amsmath}	
\usepackage{amssymb}	
\usepackage{newtxtext,newtxmath}
\usepackage[normalem]{ulem}

\newcommand{\HI}{H\textsc{\scriptsize I}}
\newcommand{\HII}{H\textsc{\scriptsize II}} 
\newcommand{\HeII}{He\textsc{\scriptsize II}}
\newcommand{\OII}{O\,\textsc{\scriptsize II}}
\newcommand{\HeIII}{He\textsc{\scriptsize III}}

\newcommand*\colourcheck[1]{\expandafter\newcommand\csname #1check\endcsname{\textcolor{#1}{\ding{52}}}}
\colourcheck{blue}
\colourcheck{green}
\colourcheck{red}
\def\app#1#2{%
  \mathrel{%
    \setbox0=\hbox{$#1\sim$}%
    \setbox2=\hbox{%
      \rlap{\hbox{$#1\propto$}}%
      \lower1.1\ht0\box0%
    }%
    \raise0.25\ht2\box2%
  }%
}

\usepackage[T1]{fontenc}
\DeclareRobustCommand{\VAN}[3]{#2}
\let\VANthebibliography\thebibliography
\def\thebibliography{\DeclareRobustCommand{\VAN}[3]{##3}\VANthebibliography}
\usepackage{graphicx}	
\usepackage{hyperref}
\usepackage{color}
\usepackage{multirow}
\usepackage{verbatim}
\usepackage{tabularx} 
\usepackage{xspace}	
\usepackage{xcolor}
\usepackage{sistyle}
\usepackage{orcidlink}
\SIthousandsep{,}
\graphicspath{{./}{figures/}}

\defcitealias{bailer-jones:21}{BJ21}
\DeclareUnicodeCharacter{2212}{-}
\title[Cosmic web Lyman-$\alpha$ glow]{The cosmic web's Lyman-$\alpha$ glow at $z \approx 2.5$; hydrodynamic models, dust, and wide-field, narrow-band detection}
\author[Sokoliuk et al.]{Oleksii Sokoliuk\thanks{E-mail: oleksii.sokoliuk@mao.kiev.ua}$^{1,2,3}$, 
John K. Webb\thanks{E-mail: jw978@cam.ac.uk}$^{4,5,6}$, 
Kenneth M. Lanzetta$^7$, 
Michael M. Shara$^8$, 
Stefan Gromoll$^{9}$, 
\newauthor
James S. Bolton$^{10}$, 
Robert F. Carswell$^4$,
Gaspar Galaz$^{11}$, 
C\'edric Ledoux$^{12}$, 
Gaspare Lo Curto$^{12}$, 
\newauthor
Alain Smette$^{12}$, 
David Valls-Gabaud$^{13}$, 
Anja von der Linden$^7$, 
Frederick M. Walter$^7$, 
Joris Witstok$^{14,15}$. \\ \\ 
$^1$ Main Astronomical Observatory of the National Academy of Sciences of Ukraine, 27 Akademik Zabolotny St., Kyiv, 03143, Ukraine\\
$^2$ Astronomical Observatory, Taras Shevchenko National University of Kyiv, 3 Observatorna St., 04053 Kyiv, Ukraine\\
$^3$ Department of Physics, University of Aberdeen, Aberdeen AB24 3UE, UK \\
$^4$ Department of Physics, Faculty of Arts and Sciences, Beijing Normal University, Zhuhai 519087, China\\
$^5$ Institute of Astronomy, University of Cambridge, Madingley Road, Cambridge CB3 0HA, UK \\
$^6$ Big Questions Institute, Level 4, 55 Holt St., Surry Hills, Sydney, NSW 2010, Australia \\
$^7$ Department of Physics and Astronomy, Stony Brook University, Stony Brook, NY 11794-3800, USA \\
$^8$ Department of Astrophysics, American Museum of Natural History, New York, NY 10024, USA \\
$^{9}$ Amazon Web Services, 410 Terry Ave. N, Seattle, WA 98109, USA \\
$^{10}$ School of Physics and Astronomy, The University of Nottingham, University Park, Nottingham, NG7 2RD, UK \\
$^{11}$ Instituto de Astrofísica, Pontificia Universidad Católica de Chile, Vicuña Mackenna 4860, Macul, Santiago, Chile \\
$^{12}$ European Southern Observatory, Alonso de Córdova 3107, Vitacura, Santiago, Chile\\
$^{13}$ Observatoire de Paris, LERMA, CNRS UMR 8112, 61 Avenue de l'Observatoire, 75014 Paris, France \\
$^{14}$ Cosmic Dawn Center (DAWN), Copenhagen, Denmark \\
$^{15}$ Niels Bohr Institute, University of Copenhagen, Jagtvej 128, DK-2200, Copenhagen, Denmark
}
\date{Accepted xxxx. Received xxxx; in original form xxxx}
\pubyear{2025}
\begin{document}
\label{firstpage}
\pagerange{\pageref{firstpage}--\pageref{lastpage}}
\maketitle

\begin{abstract}
The diffuse Lyman-$\alpha$ glow of the cosmic web has long been predicted but has so far eluded direct detection over cosmologically significant volumes. We construct synthetic Lyman-$\alpha$ surface-brightness maps using five state-of-the-art hydrodynamic simulations (\texttt{IllustrisTNG, EAGLE, CROCODILE, SIMBA, and Sherwood}),  modeling recombination, collisional excitation, star formation, and localized dust attenuation. Our study focuses on the redshift range $2.0<z<2.7$, motivated by the numerous detailed studies of the COSMOS region. Significant variations are seen in the results obtained from these independent simulations. Using the Anderson-Darling statistic to probe these statistical differences, we demonstrate that a $5\sigma$ statistical detection of the total intergalactic and circumgalactic Lyman-$\alpha$ emission is achievable with current facilities at flux thresholds brighter than $\sim 8 \times 10^{-17} \text{ erg s}^{-1}\text{ cm}^{-2}\text{ arcsec}^{-2}$. Conversely, isolating the underlying low-density component of the cosmic web requires ultra-deep sensitivity, with the most optimistic simulation (IllustrisTNG) reaching a 5$\sigma$ detection only for background noise levels below $\sigma \sim 2 \times 10^{-19} \text{ erg s}^{-1}\text{ cm}^{-2}\text{ arcsec}^{-2}$. These quantitative limits validate the feasibility of ongoing wide-field narrow-band campaigns, opening a new era of empirical intergalactic cartography.
\end{abstract}


\section{Introduction}\label{sec:Intro}

The notion that matter in the Universe is distributed in a filamentary pattern has its roots in the pioneering theoretical work of \cite{Zeldovich1970}; see also the commentary by \cite{Shandarin2009}. The cosmic web was first visualised in the early N-body calculations of \cite{Klypin1983}. Soon afterwards, observations changed the picture dramatically, with the survey of \cite{Geller1989}, who demonstrated that the galaxy distribution did indeed follow a filamentary pattern. Subsequent theoretical work showed that such a pattern arises naturally within the $\Lambda$CDM cosmological model \citep{Efstathiou1985, Davis1985, Bond1996}. A significant fraction of the matter in the Universe is thought to  reside within this cosmic web \citep{Aragon2010,Cautun2014,Eckert2015,Burchett2020,Navdha2025}. Detailed mapping of this interconnected network of gaseous filaments, galaxies, quasars, dark matter, and voids is central to a comprehensive understanding of the origin and evolution of our Universe. Cosmic web cartography provides a direct means of testing both analytic \citep{Kitaura2024} and hydrodynamic descriptions of large-scale structure formation (as explored in the present study). Observations will distinguish between competing theoretical frameworks of dark matter, the dominant mass constituent of the Universe, including fuzzy dark matter models \citep{Dome2023}, axion-based models \citep{Rogers2023}, and alternative gravity theories \citep{CANTATA2021,Boldrini2024,Sokoliuk2025a}. Further, cosmic web measurements will elucidate galaxy formation and evolution theories \citep{Libeskind2018}, filament connectivity and relation to structure formation and cosmology \citep{Codis2018}, enabling us to understand the relation between cosmic flows and dynamics and the cosmic web \citep{Shandarin2011, Kugel2024}, and lead to a more comprehensive inventory of the matter reservoirs of the Universe \citep{Connor2024}. The broad scientific interest in the cosmic web extends even to a quantitative comparison with the human brain's neuronal network \citep{Vazza2020}.

Several studies have reported detections of luminous structures associated with the cosmic web: six Lyman-$\alpha$ emitting objects plus one damped Lyman-$\alpha$ absorber form a filamentary morphology at least 5 cMpc in extent \citep{Moller2001}; weak gravitational lensing measurements allow the inference of the existence of a dark-matter filament connecting the two main components of the Abell 222/223 supercluster system \citep{Dietrich2012}; Lyman-$\alpha$ emission around a high redshift quasar, extending beyond the virial radius of a potential dark matter halo, may be interpreted as intergalactic and part of the cosmic web in which the quasar itself resides \citep{Cantalupo2014}; eROSITA X-ray images reveal hot gaseous bridges connecting galaxies within the Abell 3391/95 triple cluster system \citep{Reiprich2021}. These studies represent valuable developments in our empirical knowledge of the cosmic web, but primarily focus on high-density peaks within a possible filamentary pattern. They do not correspond to direct detections of the more tenuous, extensive gaseous filamentary structures predicted by $\Lambda$CDM, visualised in many hydrodynamic simulations, and expected to be associated with the well-studied Lyman-$\alpha$ forest of neutral hydrogen absorption lines seen in the spectra of distant quasars.

\cite{Umehata2019} detect rest-frame ultraviolet Lyman-$\alpha$ emission from multiple filaments extending over one megaparsec between galaxies within the SSA 22 proto-cluster at a redshift of 3.1, measured at a surface brightness level $\sim 10^{-19}$ erg s$^{-1}$ cm$^{-2}$ arcsec$^{-2}$. They report star formation and AGN activity within that structure and speculate that the ionising radiation from such sources powers nearby filamentary Lyman-$\alpha$ emission. \cite{Tornotti2025ApJL} report Lyman-$\alpha$ emission measurements coming from 19 distinct regions embedded in a $\sim 5$ Mpc (comoving) section of a cosmic web filament at redshift 4.0. The 19 regions detected emit Lyman-$\alpha$ photons with surface brightnesses spanning the range $1 - 5 \times 10^{-19}$ erg s$^{-1}$ cm$^{-2}$ arcsec$^{-2}$. \cite{Tornotti2025NatAs} report the detection of cosmic web emission connecting two quasar-host galaxies at a redshift of $z \approx 3.22$ in the {\it MUSE Ultra Deep Field} \citep{Fossati2019}.

Predictions of emission from the Lyman-$\alpha$ forest were explored almost 40 years ago \citep{Hogan1987,Gould1996}. A theoretical model for the detection of Lyman-$\alpha$ emission is developed in \cite{Byrohl2023}, computing the summed emission from discrete Lyman-$\alpha$ emitters (also see \cite{Khoraminezhad2025}), halos, blobs (LABs), and cosmic web filaments. Prospects for detecting Lyman-$\alpha$ emission from the large-scale cosmic web have been revisited using both 13 nm filters \citep{Renard2021}, and broad-band \citep{Renard2024} photometric measurements, cross-correlated with Lyman-$\alpha$ absorption forest data, and most recently by \cite{Liu2025}. These recent studies indicate that survey data from the Dark Energy Spectroscopic Instrument (DESI) are close to anticipated detection thresholds. Other relevant studies include \cite{Furlanetto2003,Cantalupo2005,Silva2016,Witstok2021,Byrohl2021}.

Three recent observational studies suggest possible associations between Lyman-$\alpha$ emitting objects and HI absorption clouds at the high end of the Lyman-$\alpha$ forest column density distribution, i.e. N$_{\rm \HI} \gtrsim 10^{17}$ atoms cm$^{-2}$. The study by \cite{Bacon2021,Bacon2023} describes diffuse extended Lyman-$\alpha$ emission over the redshift range 3.1 to 4.5, tracing filaments on scales up to 4 cMpc. A subsequent claim to detect Lyman-$\alpha$ emission from objects giving rise to the Lyman-$\alpha$ forest is given in \cite{Martin2023}. More recently, a small sample of partial Lyman limit absorption systems at a redshift $z \approx 3.6$ has been reported \citep{Banerjee2024}, associated with a claimed excess of Lyman-$\alpha$ emitters which indicate a spatial morphology consistent with a filamentary pattern. 

Whilst studies such as these are beginning to probe the more extended rarefied and mostly ionised regions associated with higher column density Lyman-$\alpha$ forest lines, so far only a tiny survey volume has been explored, and direct associations between Lyman-$\alpha$ emitting objects and specific absorption systems at these column densities are tenuous. Stringent tests of the $\Lambda$CDM cosmological model will ultimately be best achieved by comparing detailed 3-dimensional cartography over substantial cosmological scales against equally detailed numerical models.

The calculations described in this paper are also motivated by the surge of interest in wide-field, low-surface-brightness surveys, including: 
the Pan-STARRS Survey Telescope Project \citep{Kaiser2010}, 
Dark Energy Spectroscopic Instrument (DESI) \citep{DESI2016PartI},
Sloan Digital Sky Survey (SDSS) \citep{SDSSV2017}, 
MESSIER Mission (Ultra-Low Surface Brightness Explorer, a proposed space mission to map the very faint Universe without refractive optics) \citep{MESSIER2019SPIE}, 
Dragonfly Telephoto Array \citep{Lokhorst2019}, noting a similar study to the present paper, in which they investigate the detectability of H$\alpha$ emission from extended gaseous haloes around low redshift galaxies, 
Dragonfly Spectral Line Mapper \citep{Chen2024}, 
Huntsman Telescope \citep{Caddy2024}, 
Vera C. Rubin Observatory \citep{Brough2024}, 
Nancy Grace Roman Space Telescope \citep{Schlieder2024},
WST Widefield Spectroscopic Telescope (WST) \citep{WST2024Concept},
Euclid Space Telescope (ESA’s mission for cosmology and dark energy, launched 2023) \citep{Euclid2025Overview},
Keck Telescope Wide-Field Imager (KWFI) \citep{KWFI2022},
and others.

A specific driver for the theoretical calculations in this paper  arises from our own work using the Condor Array Telescope in New Mexico \citep{Lanzetta2023, Lanzetta2025}. Using this facility, we are carrying out deep, wide-field imaging covering most of the COSMOS field \citep{Cucciati2018}, using a narrow-band filter, complemented by luminance filter imaging. The narrow-band filter is centred at 422.5 nm with a bandpass of 1 nm, probing the redshift shell $z=2.4754 \pm 0.0030$ for Lyman-$\alpha$, corresponding to a velocity interval of 710 km\,s$^{-1}$. This image size is $\approx 2.8 \times 1.3$ degrees at a pixel scale of 0.85 arcsec pix$^{-1}$. The cosmological volume probed is around 279,000 comoving Mpc$^3$, around 4-5 orders of magnitude larger than a typical galaxy cluster. Ultra-deep wide-field imaging data of this kind is likely to lead to significant advances in understanding the physics of cosmic web formation and evolution.

In Section \ref{sec:Methods}, we describe the methodology adopted in this paper. Section 3 provides several results emerging from the simulation post-processing calculations, including the final surface brightness maps and the inferences that can be made from them. The theoretical calculations described in Section \ref{sec:dust} explore two treatments for high mass particles, optical depth thresholding and dust modelling. The first effectively rejects high column density gas such that, effectively, only diffuse, tenuous gas contributes to the statistical Lyman-$\alpha$ emission. The second, less conservative modelling choice, retains {\it all} simulation particles but applies dust attenuation based on a Galactic model, modified to allow for particle metallicities, as provided in the simulation output data. This second approach reveals interesting (but quite diverse) trends from the different cosmological simulations used in our work.

Section \ref{sec:Conclusions} summarises the main findings of this work. 

\section{Methods}\label{sec:Methods}

\subsection{Five hydrodynamic simulations}\label{sec:hydrosims}

To explore consistency amongst simulations, we have computed Lyman-$\alpha$ emission surface brightness maps using five high-resolution simulations, listed in Table \ref{tab:5sims}. The reason for using multiple cosmological simulations in this work is to provide interesting consistency/comparative checks, since each simulation treats star and galaxy formation processes differently (e.g. the number density threshold at which star formation commences within each simulation particle is assumed to be metallicity dependent in EAGLE, but IllustrisTNG takes a constant value of $n_{\rm H}\sim 0.1\;\rm cm ^{-3}$). IllustrisTNG, SIMBA, EAGLE, CROCODILE are all ``full-physics'' or ``reference physics'' simulations. We have used the ``quick-Ly$\alpha$'' version of Sherwood, in which very dense gas (overdensities $\delta \rho_b/\rho_{b} >1000$ and temperatures $T<10^5$ K) is ignored in order to speed up the calculation and to focus on the low column density gas. The Sherwood simulation data we have used is the same one used in \cite{Witstok2021}. Doing so provides an interesting comparison against the four reference physics simulations. Table \ref{tab:5sims} lists simulations at different redshifts: IllustrisTNG: $z=2.00$; SIMBA: $z=2.46$; EAGLE: $z=2.48$; Sherwood: $z=2.40$; CROCODILE: $z=2.74$. The observational comparison is with the COSMOS field is at $z\simeq 2.475$. EAGLE and SIMBA are close; Sherwood is acceptable; but IllustrisTNG at $z=2.00$ and CROCODILE at $z=2.74$ are significantly displaced. The UV background, halo mass function, thermal state, SFR density, Lyman-$\alpha$ emissivity, and cosmological surface-brightness dimming evolve over this interval. These variations were unavoidable given simulation snapshot availability, and they will clearly contribute to the spread in results illustrated later in this paper in e.g. Fig.\,\ref{fig:Asquared}.

The fundamental outputs from each simulation are, for each particle: baryonic mass, temperature, elemental abundances, and velocity vector. For all simulations where this information is not already present in the snapshots, we recompute the neutral, ionised and molecular hydrogen components in post-processing (see Sections \ref{sec:decoupling} to \ref{sec:electron_number_density}).

To compute the Lyman-$\alpha$ surface brightness distribution arising in filamentary structures, one could apply Monte-Carlo Radiative Transfer (MCRT) procedures to the simulation output data. MCRT has the appeal of including the relevant detailed physics (compared to a simpler semi-analytic method). Moreover, there are several codes available for this purpose, e.g., \texttt{COLT} \citep{Smith2015}, \texttt{RASCAS} \citep{Michel-Dansac2020}, \texttt{ART}$^2$ \citep{Li2020}, \texttt{MoCaLaTA} \citep{Laursen2009a} and its dust-attenuated version \citep{Laursen2009b}. However, observational constraints for dust albedo and opacity at $z \sim 2.5$, on scales corresponding to simulation particle masses, are not available. Therefore, the only option is to adopt simple models derived from the local Universe, diluting the accuracy benefits of the more detailed physics provided by an MCRT approach. In the present work, we are focused on the detectability of Lyman-$\alpha$ emission from the cosmic web. It has been shown that the predicted surface brightness probability distribution functions (PDFs) obtained using radiative transfer and without doing so (semi-empirical) do not differ significantly \citep{Byrohl2023}. See also the discussion on this point given in \cite{Witstok2021}. In this work, we therefore employ semi-analytic methods.

Although full radiative transfer (when dust is not included) has only a modest effect on the overall Lyman-$\alpha$ surface-brightness probability distribution \citep{Byrohl2023}, dust attenuation remains important for the highest density regions, where repeated resonant scattering substantially increases photon path lengths and hence the probability of absorption. Since these bright regions dominate the upper tail of the surface brightness distribution, whereas diffuse intergalactic emission dominates the faint end relevant to the present study, a simplified dust treatment captures the principal observational consequences without requiring computationally expensive radiative transfer calculations.

We have kept to the original cosmological parameters used for the initial conditions in generating each simulation, i.e. values for $\Omega_{\rm m}$ and $h=H_0/100\rm \; km \;s^{-1}\;Mpc^{-1}$ (noting that $\Omega_\Lambda = 1- \Omega_{\rm m}$, see Table \ref{tab:5sims}). Most simulations have kept the values of $H_0$ and $\Omega_{\rm m0}$ very close to recent cosmological measurements, e.g., from Planck. Such small variations are unlikely to dominate variations in the results seen for each simulation.

\begin{table*}
\centering
\caption{Details for each simulation used in this paper (Section \ref{sec:hydrosims}): (1) {IllustrisTNG} \citep{Springel2018}, (2) SIMBA \citep{Dave2019}, (3) EAGLE \citep{Crain2015}, (4) Sherwood, \citet{Bolton2017}, but using a simpler star formation physics implementation, as described in section 2.4 of \citet{Witstok2021}, and (5) {CROCODILE} \citep{Romano2022a, Romano2022b, Oku2022, Oku2024}. $z$ is the redshift at which each simulation is computed. $L_{\rm box}$ is the simulation box size in comoving megaparsecs. $N_{\rm tot}$ is the total number of particles (each particle comprising gas + dark matter). $m_{\rm DM}$ and $m_{\rm gas}$ are the dark matter and baryonic mass contributions to each particle in solar mass units. The comoving narrowband filter depth, $\Delta d$, relates to Eq. \eqref{eq:depth}. 
}
\label{tab:5sims}
\begin{tabular}{llllllllll}
\hline\hline
Simulation     & $H_0$ & $\Omega_{\rm m0}$ & $z$ & $L_{\rm box}$ & $N_{\rm tot}$ & $m_{\rm DM}$ & $m_{\rm gas}$   & $\Delta d$  \\ 
& $\rm [km\,s^{-1}\,Mpc^{-1}]$ & & & {[}cMpc{]} & &   $[M_\odot]$ & $[M_\odot]$ &  {[}cMpc{]} \\ \hline
IllustrisTNG$^{(1)}$  &67.74 & 0.3089 & 2.00  & 110.7 & $2\times 910^3$  & $5.97 \times 10^7$   & $1.12 \times 10^7$ & 12.09 \\ 
SIMBA$^{(2)}$          &68.00 & 0.3000 & 2.46  & 73.8  & $2\times 512^3$ & $9.6 \times 10^7$  & $1.8 \times 10^7$ & 9.99 \\
EAGLE$^{(3)}$          &67.77 & 0.3070 & 2.48  & 50    & $2\times 752^3$  & $9.70\times 10^6$  & $1.81\times 10^6$ &
9.84 \\
Sherwood$^{(4)}$       &67.80 & 0.3080 & 2.40  & 59    & $2\times 1024^3$ & $6.34 \times 10^6$  & $1.17\times 10^6$ & 9.36 \\
CROCODILE$^{(5)}$      &67.77 & 0.3099 & 2.74  & 72.4  & $2\times 512^3$  & $9.94\times 10^7$  & $1.86\times 10^7$ &
8.85 \\
\hline\hline
\end{tabular}
\end{table*}

\subsection{Intrinsic Lyman-$\alpha$ luminosities}

Lyman-$\alpha$ photons in galaxies and the IGM are produced by two processes, recombination and collisional excitation \citep{Osterbrock2006}. The first process occurs when a free electron is captured by an ionised hydrogen atom (\HII{}), with around a $2/3$ chance that the hydrogen atom will ultimately transit from the second excited state to the ground state, emitting a Lyman-$\alpha$ photon as a result\footnote{The $\sim 2/3$ probability comes from the branching ratio of the possible transitions starting from $n=3$ in the hydrogen atom, derived from quantum mechanical calculations of transition probabilities. In practice, we use a temperature-dependent value, as given in \cite{Dijkstra2014} ($f_{\rm rec,B}(T)$ in Eq. \ref{eq:1})}. Collisional excitation occurs between a neutral hydrogen atom and a free electron. If the free electron transfers sufficient kinetic energy to an \HI{} atom, its energy level changes, resulting in the emission of the Lyman-$\alpha$ photon. Both processes contribute to the observed luminosity density \citep[e.g.][]{Dijkstra2014, Silva2016},
\begin{equation}
\begin{gathered}
    \epsilon_{\rm coll} = \gamma_{\rm 1s2p}(T)n_en_{\rm \HI}E_{\rm Ly\alpha},\hfill\\
    \epsilon_{\rm rec} = f_{\rm rec,B}(T)n_en_{\rm \HII}\alpha_{\rm B}(T)E_{\rm Ly\alpha}.\hfill
    \label{eq:1}
\end{gathered}
\end{equation}
We shall see later that high column density gas ($ N_{\rm{\HI}} \gtrsim 10^{18}\;\rm cm^{-2}$) dominates the surface brightness emission for $\mathcal{S} \lesssim 10^{-18} \; \rm erg \; s^{-1}\; cm^{-2} \; arcsec^{-2}$ (see right hand column of Fig. \ref{fig:nH-T_SB-NHI} and Sections \ref{sec:T-lognH} to \ref{sec:insights}). Observational detection thresholds at present are $\mathcal{S} \sim 10^{-19} \; \rm erg \; s^{-1}\; cm^{-2} \; arcsec^{-2}$ \citep{Martin2023}. Therefore, following \cite{Witstok2021}, we use Case B throughout. For the purposes of clarification, we note that Eq. \eqref{eq:eta} uses the Case A recombination coefficient to estimate the neutral fraction in optically thin gas. In contrast, when computing Ly$\alpha$ emissivities and surface brightness, we adopt Case B, since recombinations to the ground state do not contribute to observable Ly$\alpha$ emission. For recombination, $f_{\rm rec,B}(T)$ represents the fraction of recombinations that result in Lyman-$\alpha$ emission, incorporating the effects of cascade transitions from levels $n>2$ to the $2p \rightarrow 1s$ transition. Together with the fitted Case B recombination coefficients $\alpha_{\rm B}(T)$ \citep{Cantalupo2008, Draine2011book, Dijkstra2014}, this provides a temperature-dependent effective emissivity for Lyman-$\alpha$ photons produced via recombination. $E_{\rm Ly\alpha}=1.634\times 10^{-18}$\,J is the energy of a Lyman-$\alpha$ photon. For collisional excitation, the coefficient $\gamma_{\rm 1s2p}(T)$ denotes an effective excitation rate coefficient, fitted \citep{Scholz1990,Scholz1991} as an exponential function divided into three temperature regimes ranging from $T\sim 10^3$\,K up to $T\sim 10^8$\,K \citep{Scholz1990,Scholz1991}. Although written as $\gamma_{\rm 1s2p}$, it implicitly includes excitations into higher levels ($n>2$) and the corresponding cascade branching ratios that ultimately contribute to Lyman-$\alpha$ production.

The electron number density is \citep{Katz1996,Junhan2022}
\begin{equation}
    n_e = n_{\rm \HII}+n_{\rm \HeII}+2n_{\rm \HeIII}=\frac{\rho_{\rm gas}}{\mu_e m_{ \rm gas}}. \label{eq:n_e}
\end{equation}
The value of the mean molecular weight per electron, $\mu_e$, depends on the environment (e.g whether the particle is located in the IGM or the Interstellar Medium), and on the ionisation fractions of hydrogen and helium (see Eq. \eqref{eq:mu_e}). Ions from metals are ignored, as their contribution towards the total mean molecular weight in the diffuse IGM is expected to be small.

In addition to collisional excitation, neutral hydrogen within a galaxy can also be photoionised by stellar UV, at an estimated rate of $\dot{N}_{\rm ion}\sim 10^{51}\;\rm s^{-1}\,Mpc^{-3}$ in the redshift range $z\sim 2-6$ \citep{Gaikwad2023}. One approach would be to calculate the ionisation rate for each star-forming particle using a code like \texttt{BPASS} \citep{2017PASA...34...58E} or \texttt{STARBURST99} \citep{1999ApJS..123....3L} and then adopt the approach of \cite{Byrohl2023}. However, as discussed in \cite{Byrohl2023}, young stellar populations may be unresolved in simulations having lower resolution than that of TNG50, leading to poorly estimated star formation Lyman-$\alpha$ emissivity $\epsilon_{\rm sf}$. Since the resolution of the five simulations used in this paper is lower than that of TNG50, we implement a simpler model, and assume that star formation emissivity is directly proportional to the star formation rate for a particle $\dot{M}_\star$ \citep{Byrohl2021},
\begin{equation}
    \epsilon_{\rm sf}=10^{42}\bigg(\frac{\dot{M}_\star}{M_\odot \rm yr^{-1}}\bigg)\frac{\rm erg \;s^{-1}}{V_{\rm cell}}.
\end{equation}

The gas density, 3D particle velocity, \HI{}/\HII{} number densities, star formation rate, metallicities and Lyman-$\alpha$ emissivities are derived for each particle in each hydrodynamic simulation and then placed onto a Cartesian 3-dimensional grid using ParticleGridMapper.jl \citep{web:particlegridmapper} and SPHtoGrid.jl codes \citep{web:sphtogrid}. The cell size is variable and depends on the simulation (see Table \ref{tab:5sims}) and the desired resolution of the final surface brightness image. It can be expressed as $V_{\rm cell}=(L_{\rm box}/\rm N_{\rm pix})^3$, where $N_{\rm pix}$ is chosen such that the cell size is considerably bigger than the inter-particle separation to prevent any interpolation artefacts from appearing.
Then, for each cell (i.e. one element in the 3-dimensional grid) with a corresponding volume $V_{\rm cell}$, we calculate the Lyman-$\alpha$ luminosity using 
\begin{equation}
    L_{\mathrm{Ly}\alpha,\mathrm{cell}}=(\epsilon_{\mathrm{rec}}+\epsilon_{\mathrm{coll}}+\epsilon_{\rm sf})V_{\rm cell}. \label{eq:Llya}
\end{equation}
The exception is Sherwood, where we do not include $\epsilon_{\rm sf}$ since the ``star formation rate'' values provided  in the Sherwood model are not physical (see comment in Section \ref{sec:hydrosims}) and would over-produce star particles if used. Different methods for estimating the Lyman-$\alpha$ intrinsic luminosity, including self-shielding, varying star formation and AGN feedback effects, are discussed in \cite{Faucher2010}.

To model narrowband imaging, specifically for a 1 nm waveband filter centred on the Lyman-$\alpha$ line at the redshift of our snapshot (Section \ref{sec:addnoise}), we consider a slice of comoving size:
\begin{equation}
    \Delta d = c\int ^{z+\Delta z}_{z}\frac{dz'}{H(z')},
    \label{eq:depth}
\end{equation}
where $\Delta\lambda=\rm 1\;nm$ and $\Delta z =\Delta \lambda /\lambda_{\rm Ly\alpha}$, where $\lambda_{\rm Ly\alpha}=121.57\;\rm nm$ is the rest-frame wavelength of Lyman-$\alpha$ emission. All particles that do not fit within a slice $\Delta d$ are ignored, as are particles having a projected line-of-sight velocity falling outside the narrow band filter range. In practice, we used six slices of width $\Delta d$, each slice residing at the box edge, thereby improving the statistical sample (by a factor of six). Particles falling just outside the box edges are thus not taken into account, so our surface brightness predictions will be very slightly biased towards lower values (and our detectability estimation will therefore be slightly conservative in this sense).

\subsection{Decoupling neutral, molecular and ionised hydrogen}\label{sec:decoupling}

The simulations we consider in this work do not all keep track of the relative abundances of the ionised and molecular species in the publicly available snapshots. For EAGLE, IllustrisTNG and Sherwood we therefore compute the abundances of \HI{}, \HII{}, $\rm H_2$ and free electrons in post-processing under the assumption of photo-ionisation equilibrium with a correction for self-shielding, as we describe below. For CROCODILE and SIMBA we instead use the snapshot data. To separate hydrogen abundances, one could use a radiative transfer method \citep[e.g][]{Bauer2015}. However, for the reasons discussed in Section \ref{sec:hydrosims}, and also because considerable computing resources are required for large cosmological volumes, we instead use a semi-analytical approach \citep{Crain2017, Lagos2015}.

The atomic neutral hydrogen gas fraction, $\eta=n_{\rm \HI}/n_{\rm H}$, depends on the gas temperature, density, and photoionisation rate. We calculate $\eta$ for each particle in the simulation using fitting functions derived from a full radiative transfer treatment \citep{Rahmati2013a}. Calculations of the \HI{} fraction are made on a particle-by-particle basis prior to interpolation onto a Cartesian grid of size $N=1024^3$. Importantly, the fitting functions include the effects of self-shielding; when the particle density is sufficiently high for self-shielding to become significant, the escape fraction of Lyman-$\alpha$ photons is attenuated if dust is present, which in turn has a significant impact on the Lyman-$\alpha$ surface brightness distribution \citep{Witstok2021}. A convenient expression allowing us to compute $\eta$, given $n_{\rm H}$ and T, is \citep{Rahmati2013a}
\begin{equation}
    \eta = \frac{n_{\rm\HI{}}}{n_{\rm H}}=  \frac{B-\sqrt{B^2-4AC}}{2A},
    \label{eq:eta}
\end{equation}
where $A=\alpha_{\rm A}+\Lambda_{\rm T}$, $B=2\alpha_{\rm A}+\Gamma_{\rm phot}/n_{\rm H}+\Lambda_{\rm T}$ and $C=\alpha_{\rm A}$ with
\begin{equation}
    \Lambda_{\rm T} = 1.17\times 10^{-10} \left( \frac{T^{1/2}\exp(-157809/T)}{1+\sqrt{T/10^5}} \right),
\end{equation}
\citep{Theuns1998}, where $\Lambda_{\rm T}$ is in cm$^3$\,s$^{-1}$. Note that Eq. \eqref{eq:eta} makes a simplifying assumption since its derivation ignores the helium contribution to $n_e$. However, the radiative transfer calculations of \cite{Rahmati2013a} show this to be a small effect (also see \cite{Faucher2010,McQuin2010,Altay2011}). The ratio of the total photoionisation rate $\Gamma_{\rm phot}$ to the UV background (UVB) photoionisation rate $\Gamma_{\rm UVB}$ relation can be modelled as
\begin{equation}
    \frac{\Gamma_{\rm phot}}{\Gamma_{\rm UVB}} = (1-f)\bigg[1+\bigg(\frac{n_{\rm H}}{n_0}\bigg)^\kappa\bigg]^{\alpha_1}+f\bigg[1+\frac{n_{\rm H}}{n_0}\bigg]^{\alpha_2}. \label{eq:Gammaratio}
\end{equation}
The free parameters in Eq. \eqref{eq:Gammaratio}, i.e. $\alpha_1$, $\alpha_2$, $f$,  
$\kappa$, and the characteristic hydrogen number density $n_0$ , have been fitted to a set of cosmological simulations with various box sizes to derive best-fit parameter values applicable to $z \gtrsim 2$, appropriate for our study \citep{Rahmati2013a}. The best fit parameters are $\alpha_1 = -2.28 \pm 0.31$, $\alpha_2 = -0.84 \pm 0.11$, $f = 0.02 \pm 0.01$, $\kappa = 1.64\pm0.19$, and $n_0 = (1.003\pm0.005)\times n_{\rm H,SSh}$. The redshift dependent hydrogen number density threshold $n_{\rm H,SSh}$ is recovered from a look-up table \citep{Rahmati2013a}, using linear interpolation. The UVB photoionisation rate, $\Gamma_{\rm UVB}(z)$, is also obtained using look-up table values, again, with linear interpolation \citep{Haardt2012}. 

Now $n_{\rm H} = n_{\rm \HI{}} + n_{\rm \HII} + 2n_{\rm H_2}$, so the only remaining unknown in Eq. \eqref{eq:Gammaratio} is $n_{\rm H_2}$. In the fitting function for $\eta$, Eq. \eqref{eq:eta}, it is assumed that hydrogen comprises only neutral and ionised atoms. We wish to be as precise as possible and account for the small fraction of H$_2$ in $\rm H$ that appears at high column densities. 
This requires a slight modification of the relations introduced in this section.

\subsection{H$_2$ formation}

At column densities $N_{\rm \HI{}}\gtrsim10^{21}\rm \; cm^{-2}$, the influence of molecular hydrogen becomes significant. Approximately 97\% of the H$_{2}$ mass density is contained in these high column density absorption systems \citep[e.g][]{Zwaan2006}, for which
the H$_2$ fraction is \citep{Gnedin2011} 
\begin{equation}
    f_{\rm H_2} = \frac{\Sigma_{\rm H_2}}{\Sigma_{\rm H}} \approx \bigg(1+\frac{\Sigma_{\rm c}}{\Sigma_{\rm \HI{}+H_2}}\bigg)^{-2},
\end{equation}
where $\Sigma_{\rm H} = \Sigma_{\rm H_2} + \Sigma_{\rm \HI{}}$
and 
\begin{equation}
    \Sigma_{\rm c} = 20M_\odot \mathrm{pc}^{-2}\frac{\Psi^{4/7}(D_{\rm \scriptscriptstyle MW}, U_{\rm \scriptscriptstyle MW})}{D_{\rm \scriptscriptstyle MW}}\frac{1}{\sqrt{1+U_{\rm \scriptscriptstyle MW}D^2_{\rm \scriptscriptstyle MW}}}.
    \label{eq:sigmac}
\end{equation}

In Eq. \eqref{eq:sigmac}, $D_{\rm \scriptscriptstyle MW}\equiv Z/Z_\odot$ is the dust-to-gas mass ratio relative to the Milky Way value and $U_{\rm \scriptscriptstyle MW}$ is the normalised flux in
units of the \cite{Habing1968} radiation field, expressed as $U_{\rm\scriptscriptstyle MW}=\Sigma_{\rm SFR}/10^{-3}M_\odot \rm yr^{-1}kpc^{-2}$. The remaining unknowns are the function $\Psi$ (see \cite{Gnedin2011} and the Appendix of \cite{Lagos2015}), and the neutral hydrogen surface density, which is derived in terms of Jeans length and hydrogen density, $\Sigma_{\rm \HI{}+H_2} = \eta \rho_{\rm H}\lambda_{\rm J}$. We derive the SFR surface density directly from the SFR density, in a similar manner as we did for the neutral hydrogen surface density, $\Sigma_{\rm SFR} = \rho_{\rm SFR}\lambda_{\rm J}= \rho_{\rm SFR} c_s/\sqrt{G\rho_{\rm SFR}}$. The effective speed of sound $c_s$ for each particle is given as a function of pressure and density in \cite{Schaye2008}. A Python implementation to compute the neutral and ionised hydrogen fractions has been adapted from two existing codes \citep{Stevens2019, Witstok2021}, both of which are based on the methods described in \cite{Rahmati2013a,Gnedin2011}.
The molecular hydrogen content in the more tenuous gaseous regions is very small, most of it being concentrated within galactic halos, but our calculations allow for it in all locations, irrespective of surface density (with a very small effect on the final results).

\subsection{Deriving the electron number density}\label{sec:electron_number_density}

Only one of the simulations we use (EAGLE) does not provide the electron abundance. For that simulation 
we therefore calculate $n_e$ semi-analytically as another post-processing step, as has been done in other works \citep[e.g.][]{Lim2018}, taking $n_e = \rho_{\rm ion}[(1+f_{\rm H})/2m_p]$ where $\rho_{\rm ion}$ is the ionised gas density and $f_{\rm H} \approx 0.76$ is the mass fraction of hydrogen (the number varies very slightly for each simulation). Since $\rho_{\rm ion}/\rho_{\rm gas}\approx 1$, up to relatively high overdensities of $\log_{10} (\rho_{\rm gas}/\overline{\rho}_{\rm gas})\sim 2$, we assume that the gas is fully ionised.

For the IllustrisTNG, SIMBA, CROCODILE cases, the electron abundance field is known, but is unreliable for gas cells with $\rm SFR>0$, because in that case the abundance provided in the simulation output is just an average of both cold and hot gas phases. In order to calculate $x_e$ and subsequently $n_e$ correctly, we follow earlier work and separate cold and hot gas on a subgrid level (as the resolution of our interpolation grid is not sufficient to properly resolve hot and cold gas phases). The hot gas is assumed to be completely ionised, and the cold gas to be fully neutral. The cold gas fraction is
\begin{equation}
    x = \frac{u_h-u}{u_h - u_c},
\end{equation}
\citep{Springel2003}, where $u$ is the total specific internal energy of a particle, 
\begin{equation}
    u = \frac{k_{\rm B}T}{\mu m_{\rm H}(\gamma-1)}
\end{equation}
calculated assuming an ideal gas and an adiabatic equation of state with index $\gamma = 5/3$, and the subscripts $c$ and $h$ indicate cold and hot. For a cold gas phase, we assume that the temperature is $T_c\sim 10^3\rm \;K$, which, under the assumption of an ideal gas, can easily be translated to the internal energy $u_c$. The hot gas phase internal energy is, again following \cite{Springel2003}, as
\begin{equation}
    u_h = \frac{u_{\rm SN}}{A+1}+u_c,
\end{equation}
where $u_{\rm SN}$ is the internal energy due to supernovae feedback, defined in terms of the ``supernova temperature'' $T_{\rm SN}\sim 10^8\rm K$ and $A=100$. It is then straightforward to derive the electron abundance, since the mean molecular weight for fully ionised gas is
\begin{equation}
    \mu=(f_{\rm H} + f_{\rm He}/2 + Z/2)^{-1}.
    \label{eq:mu_e}
\end{equation}
Note that for Sherwood, there are no hot and cold phases as there is no multiphase star formation model, and it does not have a metallicity model. Thus, we do not apply the procedure above to that simulation.

\subsection{The \HI{} column density distribution function}
\begin{figure*}
    \centering
    \includegraphics[width=\linewidth]{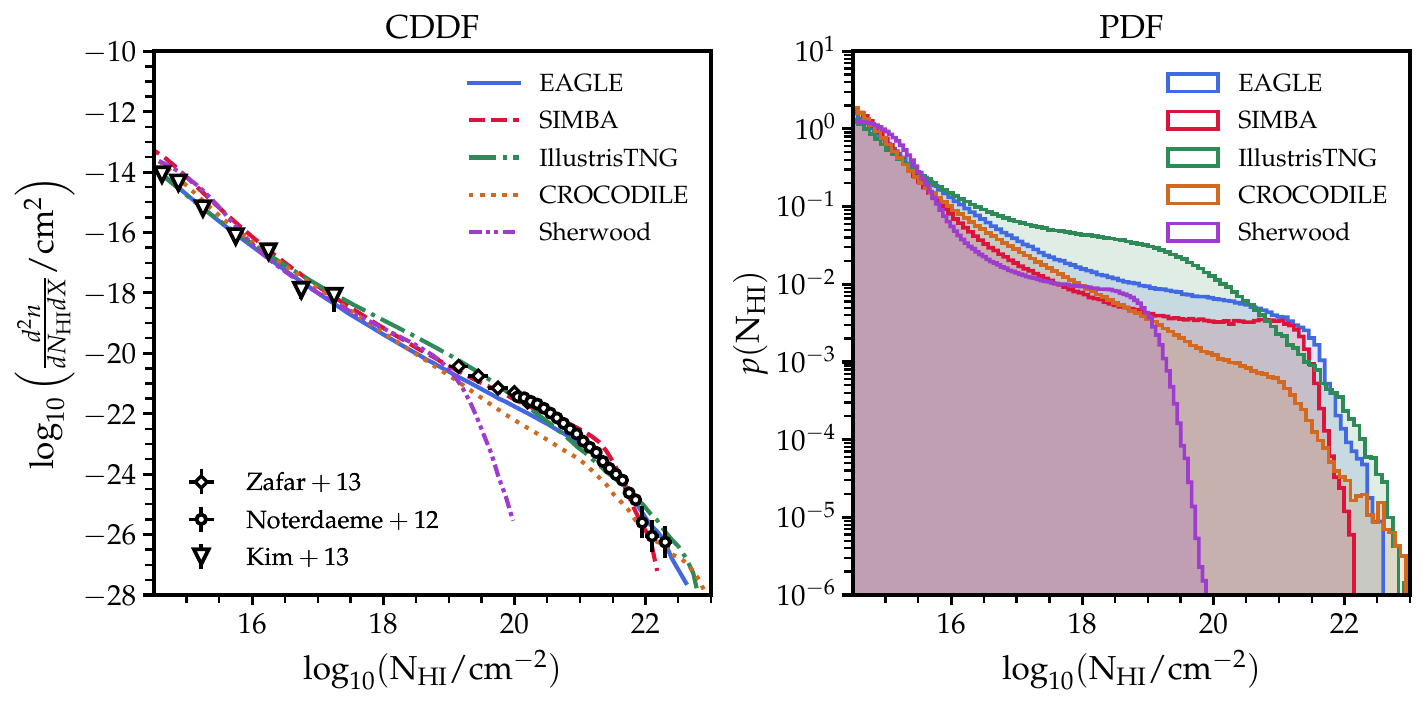}
    \caption{Comoving \HI{} Column Density Distribution Function and its probability distribution for each simulation.}
    \label{fig:CDDF}
\end{figure*}

The neutral hydrogen column density distribution function (CDDF) for each simulation is computed as follows. The CDDF can be parameterised as in e.g. \cite{Rauch1998, Rahmati2013a},
\begin{equation}
    f(N_{\rm \HI{}},X) = \frac{d^2\mathcal{N}}{dN_{\rm \HI{}}dX}
\end{equation}
where $X$ is the redshift dependent cosmological absorption length \citep{Bahcall1969}. The simulation box is divided into $m$ three-dimensional slabs, such that the absorption length can be directly related to the comoving depth of each slab, 
\begin{equation}
dX = \frac{H_0}{c}(1+z)^2 \, dL, 
\end{equation} 
with $dL \approx L_{\rm box}/m$. Following \cite{web:sphtogrid}, the column densities for each simulation are derived by integrating the number density for each particle along the line of sight within a given slab and projecting the result onto a two-dimensional grid. The CDDF is then the histogram of column densities, summed over all slabs, divided by the bin size $dN_{\rm \HI}$ and by the number of pixels over which the PDF was calculated. The slab depth $dL$ is taken to be $\approx$10 Mpc and the number of pixels for the two-dimensional grid is set at $35000 \times 35000$, such that each pixel is $\sim$kiloparsec scale.
The expected Lyman-$\alpha$ cloud size is large compared to the pixel size in the plane of the sky, but far smaller than the slab depth. The sampling used may therefore seem inappropriate. Nevertheless, in practice it seems that slab size variations of 300 $h^{-1}$kpc up to 800 $h^{-1}$kpc create a maximum of only 15\% variation in the inferred CDDF \citep{Tillman2023b}.

We briefly mention, for completeness, the alternative to the slab method: compute synthetic spectra for each simulation and then fit Voigt profiles to absorption features to extract the neutral hydrogen CDDFs. Analyses of this kind have been explored in \cite{Dave1997, Wadsley1997, Theuns1998, Theuns1999, Gurvich2017, Hiss2018}. Broadly, these methods provide reasonable agreement with results from the slab approach, which we thus adopted in this paper for simplicity. Nevertheless, the recent application of AI to automated spectral analysis \citep{Lee2021_aivpfit} marks a substantial improvement over earlier methods and can eliminate systematics associated with much slower interactive procedures. Moreover,  the quality (and quantity) of high-resolution quasar spectra has also significantly increased. More sophisticated, detailed studies are therefore now possible (e.g., automated Lyman forest modelling, fitting not just Lyman-$\alpha$ but higher-order lines) and will enable a check on systematics potentially present in analyses of both observational and simulated data.

\subsection{Escape of Lyman-$\alpha$ photons from an optically thick medium; the impact of dust.}\label{sec:dust}

For some simulation particles, the baryon number density and hence optical depth may be sufficiently high that dust can attenuate the escaping Lyman-$\alpha$ photon flux. These high-density particles contribute to the bright end of the Lyman-$\alpha$ flux distribution. Lyman-$\alpha$ photons undergo resonant scattering with neutral hydrogen atoms, causing them to follow very long, random paths before escaping a gas cloud. Since Lyman-$\alpha$ photons then scatter many times, their path length is greatly increased, enhancing the probability of encountering dust. Dust absorption (more than {\HI} photon scattering alone) is particularly effective at attenuating Lyman-$\alpha$ photon escape (see e.g. \cite{Hayes2011} who gives an empirical anti-correlation between dust attenuation and Lyman-$\alpha$ escape from galaxies).

Allowing for dust in a physically realistic way is not straightforward because we have no direct information (from the simulation output) as to the dust content or dust properties within each simulation particle. For this reason, some previous analyses simplify the situation by excluding high-density particles \citep[e.g.][]{Schaye2001, Rahmati2013a, Witstok2021}, and/or ignoring dust entirely \citep[e.g.][]{Elias2020}. However, it is important that we do not discard high-density particles entirely in this work since it is precisely these particles that form the bright end of the Lyman-$\alpha$ flux distribution (although we do discard them in Section \ref{sec:hireject} where we specifically target diffuse Lyman forest emission).

\cite{Laursen2009b} use Monte Carlo radiative transfer calculations to compute the impact of dust on Lyman-$\alpha$ emission profiles from early galaxies. \cite{Byrohl2023} also carry out radiative transfer calculations, taking into account dust, using the TNG50 simulation. Measurements of the dust abundances and properties have been made using damped Lyman-$\alpha$ absorption systems (DLAs) in quasar spectra. Whilst there is general consensus that many DLAs exhibit metal depletion patterns caused by dust, not all DLAs show evidence for dust. Recent measurements demonstrate a correlation between the dust-to-metallicity ratio and total DLA metallicity \citep{DeCia2016} and a considerable spread in DLA metallicities \citep{Dvorkin2015}. The data output from all but one (Sherwood) hydrodynamical simulation includes metallicity information for each simulation particle. Therefore, we instead adopt a simple analytic dust model that assumes a proportionality between dust and baryons (Eq. \ref{eq:n_d}). 

The cosmological simulations used in this work provide no structural information on scales below one particle in the original hydrodynamic simulation. Therefore, necessarily, our calculations assume homogeneity within each particle in each simulation. In the real Universe, matter will be clumpy on those scales and below. However, a clumpy and dusty ISM appears more transparent to radiation (both line and continuum) compared to an equivalent homogeneous ISM of equal dust optical depth, such that Lyman-$\alpha$ photons will escape more readily \citep[e.g.][]{Duval2014}. This means that our predicted surface brightness PDFs are likely to under-estimate reality, i.e., the cosmic web may be slightly easier to detect than the results we present later.

To compute the escape fraction, we use the semi-analytic expression given in \cite{Laursen2009b}, based on the slab geometry model of \cite{Neufeld1990},
\begin{equation}
    f_{\rm esc} = \frac{1}{\cosh \left(\zeta'[(\eta')^{4/3}(a\tau_{\rm \HI})^{1/3}(1-A)\tau_d]^{0.55}\right)}, \label{eq:f_esc}
\end{equation}
where the fitting parameters are provided in \cite{Laursen2009b} and are $\zeta'=2.048$, $\eta'=0.71$, the dust albedo is $A=\sigma_{\rm a}/\sigma_{\rm d}$, and $a$ is the dimensionless damping parameter. A reasonable dust albedo in this context is $A=0.32$ \citep{2001ApJ...554..778L}, and the neutral hydrogen and dust optical depths are
\begin{equation}
    \tau_{\rm \HI{}} = n_{\rm \HI{}}r\sigma_x, \quad\tau_{\rm d} = n_{\rm d}r\sigma_{\rm d},
\end{equation}
where, since we are considering photon escape, we take $r$ measured from the slab centre (Eq. \eqref{eq:Jeans}). The neutral hydrogen number density is calculated via Eq. \eqref{eq:eta}, and the dust number density is
\begin{equation}
    n_{\rm d} = (n_{\rm \HI{}} + f_{\rm ion}n_{\rm \HII{}})\frac{\sum_i Z_i}{\sum_i Z_{i,0}}. \label{eq:n_d}
\end{equation}

Two dust cross-section models are considered in \cite[e.g.][]{Laursen2009b}: the Small Magellanic Cloud (SMC) and Large Magellanic Cloud (LMC) extinction curves \citep{Pei1992}, 
\begin{equation}
    \sigma_{\rm d}/10^{-21}\;\mathrm{cm^{2}} = \begin{cases}
        0.395 + 1.82 \times 10^{-5} (T /10^4 \mathrm{\; K})^{1/2} x \;\;\;\mathrm{for\;SMC},\\
        0.723 + 4.46 \times 10^{-5} (T /10^4\mathrm{\; K})^{1/2} x\;\;\;\mathrm{for\;LMC}. \label{eq:sigma_d}
    \end{cases}
\end{equation}
Here we adopt the SMC extinction curve, since stellar populations in the SMC are younger than the LMC \citep[e.g.][]{Yanchulova2017}, so more likely to bear similarities to the high redshift Universe. For detailed discussions on this point see \cite{Li2021, Yanchulova2017} and references therein. In principle, other dust models could be used e.g. \cite{Vogelsberger2020} study three models in the context of the IllustrisTNG simulation. Although alternative empirical extinction curves (e.g. LMC or Milky Way) could equally well have been adopted, we do not attempt a survey of such models. There is presently no observationally established dust law for diffuse cosmic-web gas at $z \approx 2.5$, so no particular local extinction curve can be regarded as preferred. More fundamentally, changing the extinction curve modifies only the adopted dust attenuation prescription entering Eq. \eqref{eq:sigma_d}, whereas changing the underlying hydrodynamical simulation simultaneously changes the gas density, temperature, metallicity, ionisation structure, star-formation rate, and velocity fields that determine both the intrinsic Lyman-$\alpha$ emissivity and the escape fraction. Our comparison of independent state-of-the-art hydrodynamical simulations already demonstrates that these differences between state-of-the-art simulations produce substantial variations in the predicted surface-brightness distributions. We therefore expect the uncertainty associated with the underlying hydrodynamical model to exceed that arising from the choice among plausible local extinction curves, and consequently use the SMC extinction curve as a physically motivated illustrative prescription rather than implying that it provides a unique description of high-redshift dust.

The neutral hydrogen cross section is
\begin{equation}
    \sigma_x = f_{12}\frac{\sqrt{\pi} q_e^2}{m_e c \Delta \nu_{\rm D}}H(a,x), \label{eq:sigma_x}
\end{equation}
where $f_{\rm 12}= 0.4162$ is the Lyman-$\alpha$ oscillator strength, $q_e$ and $m_e$ are the electron charge and mass, and $c$ is the speed of light. $H(a,x)$ is the Voigt function with $a=\Delta \nu_{\rm L}/2\Delta \nu_{\rm D}$, $x=(\nu-\nu_0)/\Delta v_{\rm D}$, $\Delta \nu_{\rm D}$ is the thermal (Doppler) line width, $\Delta \nu_{\rm L}$ is the natural (Lorentzian) line width, $\nu$ is the photon frequency, and $\nu_0=2.46607 \times 10^{15}\rm \;Hz$ is the Lyman-$\alpha$ line centre frequency. The Lorentzian Lyman-$\alpha$ line width is $\Delta \nu_{\rm L}= 9.936 \times 10^7\rm \; Hz$, the Doppler width is $\Delta \nu_{\rm D} = (v_{\rm th} /c)\nu_0$, and the thermal velocity dispersion is $v_{\rm th}=(2 k_{\rm B}T/m_{\rm H})^{1/2}$. Since the calculations described here do not use radiative transfer, we do not account for scattering modifications of the photon frequency, i.e., a scattered photon retains its original frequency. However, the particle mass within a simulation is much smaller than a typical galaxy mass (see Table \ref{tab:5sims}), and each galaxy in a simulation is thus described by thousands of simulation particles, each having a 3D velocity vector. Calculations for $f_{\rm esc}$ are made on a particle-by-particle basis, so the photon frequency variation associated with galactic dispersion is accounted for \textit{a priori}.

To estimate the cloud radius $r$, we take $r\sim  L_J/2$, i.e., a typical gas cloud's radius is approximately equal to the Jeans scale \citep{Schaye2001}, given by 
\begin{align}
    L_J &\sim 10^2\;\mathrm{kpc}\;
    \left(\frac{N_{\rm \HI{}}}{10^{14}\;\mathrm{cm^{-2}}}\right)^{-1/3}
    \left(\frac{T}{10^4\;\mathrm{K}}\right)^{0.41} \notag\\
    &\quad \times
    \left(\frac{\Gamma}{10^{-12}\;\mathrm{s^{-1}}}\right)^{-1/3}
    \left(\frac{f_g}{0.16}\right)^{2/3},
    \label{eq:Jeans}
\end{align}
where $f_g = \Omega_{\rm b}/\Omega_{\rm m} \, (\approx 0.16)$ (specific value depends on the simulation - see cosmological parameters in Table \ref{tab:5sims}). The column density of neutral hydrogen for a Lyman-$\alpha$ cloud is
\begin{align}
    N_{\rm \HI} &= 2.3 \times 10^{13}\;\mathrm{cm^{-2}}\;
    \left(\frac{n_{\rm H}}{10^{-5}\;\mathrm{cm^{-3}}}\right)^{3/2}
    \left(\frac{T}{10^4\;\mathrm{K}}\right)^{-0.26} \notag\\
    &\quad \times
    \left(\frac{\Gamma}{10^{-12}\;\mathrm{s^{-1}}}\right)^{-1}
    \left(\frac{f_g}{0.16}\right)^{1/2},
    \label{eq:NHI}
\end{align}
where $n_{\rm H}$ and $T$ are provided in the simulation data. The photoionisation rate $\Gamma$ is 
\begin{equation}
    \Gamma = \int^\infty_{\nu _ {\rm L}}\frac{4\pi J(\nu) \sigma(\nu)}{h\nu}d\nu,
\end{equation}
where $J(\nu)$ is the integrated UV background intensity, $\nu_{\rm L}$ and is the Lyman-limit frequency. We use the $\Gamma$ values tabulated in \cite{Haardt2012}. We have not explored different UV background models, but note that \cite{Bird2014} suggests the calculated CDDF (Section \ref{sec:CDDF}) is fairly insensitive to the UV background, at least for column densities $\log_{10} N_{\rm\HI} \gtrsim 17$.

\begin{figure*}
    \centering
    \includegraphics[width=0.68\linewidth]{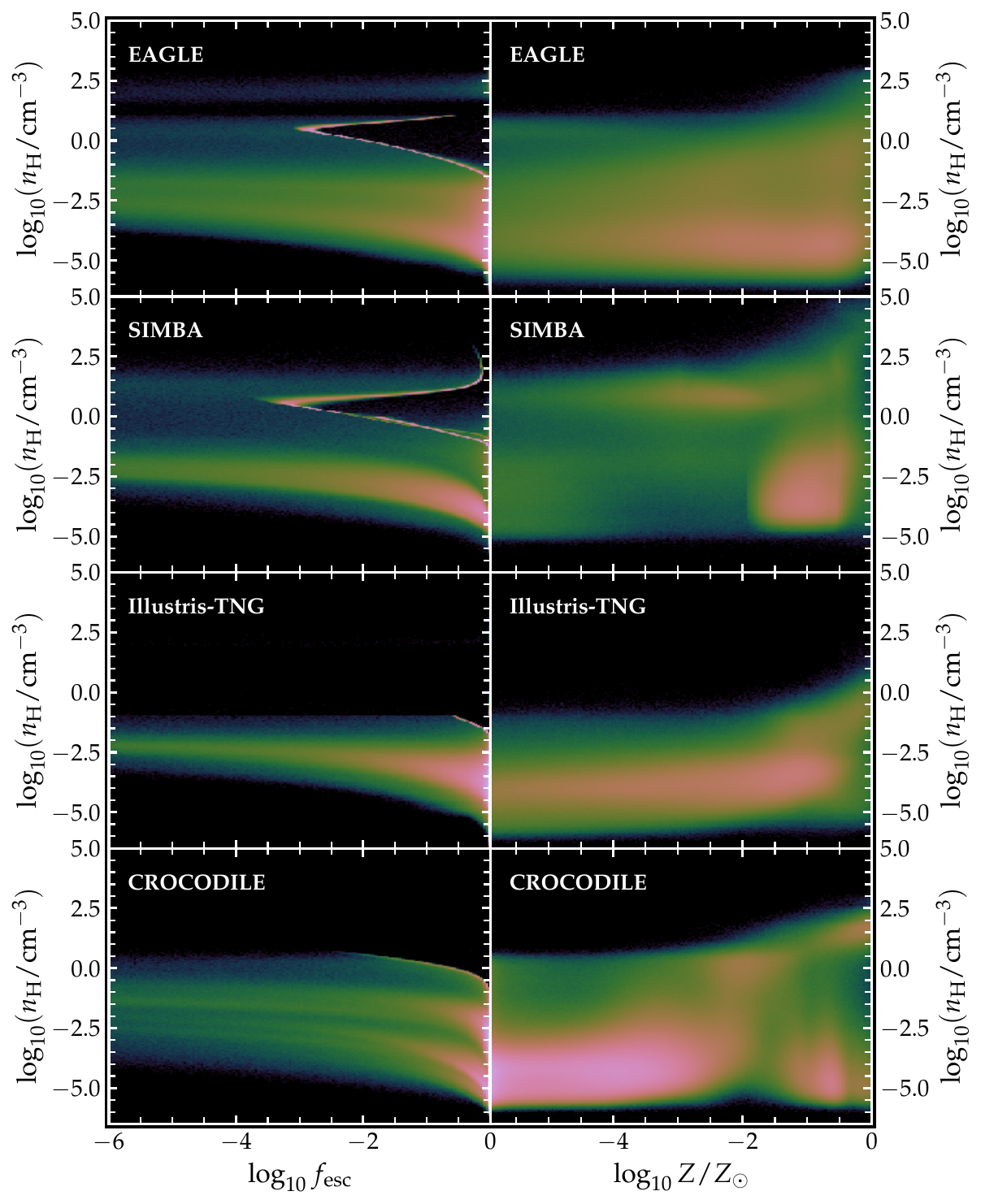}
    \caption{Lyman-$\alpha$ escape fraction and metallicity {\it vs.} hydrogen density. Sherwood does not provide metallicity per particle, so it is not shown. The wedge-like feature observed in the left column is a consequence of the artificial relation between gas density and temperature imposed in these models at high values of $\log_{10}(n_{\rm H}/{\rm cm}^{-3})$ by the sub-grid star formation prescription (as illustrated for example by the abrupt change in crossing the star-formation threshold in left column of Fig.\,\ref{fig:nH-T_SB-NHI}).
    }
    \label{fig:metal-nhi}
\end{figure*}

Fig. \ref{fig:metal-nhi} shows the metallicity-escape fraction relations for EAGLE, SIMBA, IllustrisTNG, CROCODILE (Sherwood does not provide metallicity so is not plotted). 

\subsection{Removing high density particles - hydrogen number density upper limit}\label{sec:hireject}

An alternative to the previous dust calculation is to cull high-density simulation particles where dust would be expected to play an important role. Previous studies that have applied this kind of approach include \cite{Schaye2001, Pontzen2008, Altay2011, Voort2012, Rahmati2013a, Rahmati2013b, Witstok2021}. When the neutral hydrogen column density is sufficiently high, incident UV background Lyman-$\alpha$ photons are prevented from fully penetrating the absorbing gas cloud, i.e., self-shielding ceases to be negligible once the optical depth becomes much larger than $\tau_{\rm \HI} \approx 1$. Imposing a Jeans length limit on the gas cloud size \citep{Schaye2001, Rahmati2013a}, we can obtain a hydrogen number density upper limit \citep{Zheng2002, Chardin2018}. Imposing such a limit evidently dramatically reduces the maximum predicted narrow-band Lyman-$\alpha$ surface brightness. We make use of Table 2 in \cite{Rahmati2013a}, which provides self-shielding constraints for 3 UV background models for $0<z<5$. In Section \ref{sec:Results}, we show results based on both approaches, i.e., the dust calculation approach of Section \ref{sec:dust} and culling high-density particles as described here.

\subsection{Final surface brightness maps}\label{sec:SBmaps}

The surface brightness maps are then obtained from the particle luminosities by projecting them onto a two-dimensional plane. Let each Lyman-$\alpha$ luminosity in the data cube of Eq. \eqref{eq:Llya} be $L_{ijk}$.
Indices $i,j$ denote the pixel position in the plane of the sky, and $k$ corresponds to the line-of-sight direction. The projected surface luminosity is
\begin{equation}
    \Sigma_{ij} = \frac{1}{A_{\rm patch}}\sum _k L_{ijk},
\end{equation}
where $A_{\rm patch}$ is the patch of sky area that the pixel with the integrated luminosity $\sum_k L_{ijk}$ occupies (in proper $\rm cm^{2}$). Then
\begin{equation}
    \Omega_{\rm patch} = \frac{A_{\rm patch}}{D_{\rm A}^2(z)},
\end{equation}
where $D_{\rm A}(z)$ is the angular diameter distance and $\Omega_{\rm patch}$ is in $\rm sr$. The flux observed at $z=0$ from the integrated luminosity $\sum_k L_{ijk}$ is 
\begin{equation}
    F_{\rm patch} = \frac{1}{4\pi D_{\rm L}^2}\sum_k L_{ijk},
\end{equation}
where $D_{\rm L}(z)$ is the luminosity distance and the appropriate units of $F_{\rm patch}$ are $\rm erg \; s^{-1}\; cm^{-2}$. 

Finally, the surface brightness of a particular pixel is its flux divided by the area it occupies on the sky,
\begin{equation}
\mathcal{S} = \frac{F_{\rm patch}}{\Omega_{\rm patch}}=\frac{D_{\rm A}^2(z)}{4\pi A_{\rm patch} D_{\rm L}^2(z)}\sum_k L_{ijk} = \frac{\Sigma_{ij}}{4\pi (1+z)^4},
\label{eq:S_final}
\end{equation}
where we have used $D_{\rm A}(z)/D_{\rm L}(z)=1/(1+z)^2$ and where we have converted from sr to arcseconds 
such that the units of $\mathcal{S}$ are the usual convention of $\rm erg \; s^{-1}\; cm^{-2} \; arcsec^{-2}$.

\section{Results} \label{sec:Results}

\subsection{The \HI{} column density distribution function}\label{sec:CDDF}

We show the CDDF derived from each simulation on the left-hand side of Fig. \ref{fig:CDDF}, and the corresponding PDFs are shown on the right. The observational data are also illustrated, obtained from \cite{Zafar2013AA,Noterdaeme2012AA,Kim2013AA}. All five simulations provide a good match to the observed CDDF data points up to $\rm N_{\rm \HI{}}\sim 10^{17.5}\;\mathrm{cm^{-2}}$. The lack of data points in the approximate column density range $10^{17.5}\;\mathrm{cm^{-2}} \lesssim \rm N_{\rm \HI{}} \lesssim 10^{19}\;\mathrm{cm^{-2}}$ prevents a direct check there, but for $\rm N_{\rm \HI{}} \gtrsim 10^{19}\;\mathrm{cm^{-2}}$, EAGLE, SIMBA and IllustrisTNG all fit the data well; only relatively small deviations between the models are seen, a consequence of varying resolution and the prescription used for the star formation physics. This can be seen more conspicuously in the PDFs, which exclude the normalisation used to calculate the CDDFs.

CROCODILE shows a more significant deviation in the column density range $\rm N_{\rm \HI{}}\sim 10^{19}-10^{22.3}\;\mathrm{cm^{-2}}$. The most significant discrepancy between the observational and simulation data has been observed for the Sherwood case, which cannot provide enough $\rm \HI{}$ absorbers to be consistent with the observational data for $\rm N_{\rm \HI}\gtrsim 10^{19}\;cm^{-2}$. This is caused by the simplified star and galaxy formation physics adopted; see e.g. \cite{Viel2004}.

\subsection{Physical insights from the $\log_{10} T$-$\log_{10} n_{\rm H}$ and $\log_{10}\mathcal{S}$-$\log_{10} N_{\rm \HI{}}$ phase diagrams}\label{sec:insights}

Fig. \ref{fig:nH-T_SB-NHI} and its parameterisations (Equations \ref{eq:nh-T} and \ref{eq:SB-NHI}) provide interesting insights into the thermal state of the diffuse IGM, as represented by each simulation. Although these relations exhibit complex shapes, we can nevertheless perform a few simple analyses to extract useful constraints, as discussed next.

\subsubsection{$\log_{10} T$-$\log_{10} n_{\rm H}$ relation}\label{sec:T-lognH}

Fig. \ref{fig:nH-T_SB-NHI} shows that the $T$-$n_{\rm H}$ relations for each simulation have generally similar forms, although at higher particle number densities, $n_{\rm H}\gtrsim 10^{-1}\rm \;cm^{-3}$, simulations differ in the way star formation is computed. Dense regions are reasonably rare, so assuming a simpler star formation calculation at lower particle density (where metallicities are low and star formation plays a less important role) helps to increase overall computational efficiency. The Sherwood simulation has no data at high particle number densities because (as mentioned previously, Section \ref{sec:hydrosims}) a simplified subgrid baryonic physics prescription is used, and this tends to under-produce the higher density regions. The consequence of this is also observed in the CDDF (Section \ref{sec:CDDF} and Fig. \ref{fig:CDDF}).

This low-density region of the $T$-$n_{\rm H}$ relation represents the diffuse IGM, where gas temperature is determined by both cooling and photoionisation processes \citep{Dave1999,Hernquist1996}. Avoiding very low values (near the particle mass resolution limit), the data can be modelled as a power-law,
\begin{equation}
T(n_{\rm H}) = T_0 \, \left ( \frac{n_{\rm H}}{\overline{n}_{\rm H}} \right)^{\gamma - 1} \label{eq:nh-T} 
\end{equation}
where $\gamma$ is the power-law index quantifying the thermal state of the diffuse IGM and the factor $T_0$ has units of K. The fitting was carried out by first calculating the modes for each column in the left panels of Fig. \ref{fig:nH-T_SB-NHI}, i.e. deriving the modes at constant $n_{\rm H}$. The mode was used after having tried the median, since the former gave a better visual fit. We then applied \texttt{SciPy} \citep{Virtanen2020} non-linear least squares to the set of modes, not to the images themselves, fitting to manually selected data regions (shown as dotted lines in Fig. \ref{fig:nH-T_SB-NHI}), to reduce the impact of end effects. Although the results obtained (including parameter uncertainty estimates) depend on these decisions, the resulting fits look reasonable.

The measurements show considerable variation in the value of $\gamma$ (Table \ref{tab:powerlawfits}). There are (at least) two reasons for this: redshift dependence (note the different redshifts for each simulation, given in Table \ref{tab:5sims}) and different gas cooling/heating rates amongst the simulations. Previous measurements in the literature exhibit a similarly wide spread. For example, \cite{Hui1997} find $1.4 \lesssim \gamma \lesssim 1.6$ at $z  \approx 3$ (the range corresponding to possible values of the scalefactor at reionisation). \cite{Hiss2018} (their table 4), find $\gamma = 1.45 \pm 0.08$ at $z=2.5$ (taking weighted mean of their values at $z=2.4$ and $2.6)$. \cite{Lukic2015} give $1.545 \lesssim \gamma \lesssim 1.552$ (estimated from their figure 4). Given the various assumption/models in these cases, it is difficult to draw any meaningful conclusions, other than noting the generally good agreement. 

Table \ref {tab:powerlawfits} presents the fitted slopes $\gamma$ and the normalisations $T_0$ for each simulation. These values can be compared against Lyman-$\alpha$ forest values obtained using high-resolution quasar spectra and Voigt profile modelling \citep{ascl:VPFIT2014,WebbVPFIT2021,Lee2022Addendum,web:VPFIT}. The $b$-$N_{\rm \HI}$ cut-off \citep[e.g.][]{Rorai2018} gives $T_0 = 1.56 \pm 0.44 \times 10^4$ K, and $\gamma = 0.45 \pm 0.17$, in good agreement with the values in Table \ref {tab:powerlawfits}, as expected.

\subsubsection{$\log_{10}\mathcal{S}$-$\log_{10} N_{\rm \HI{}}$ relation}\label{sec:SB-logNHI}

\begin{figure*}
\centering
\includegraphics[width=0.8\linewidth]{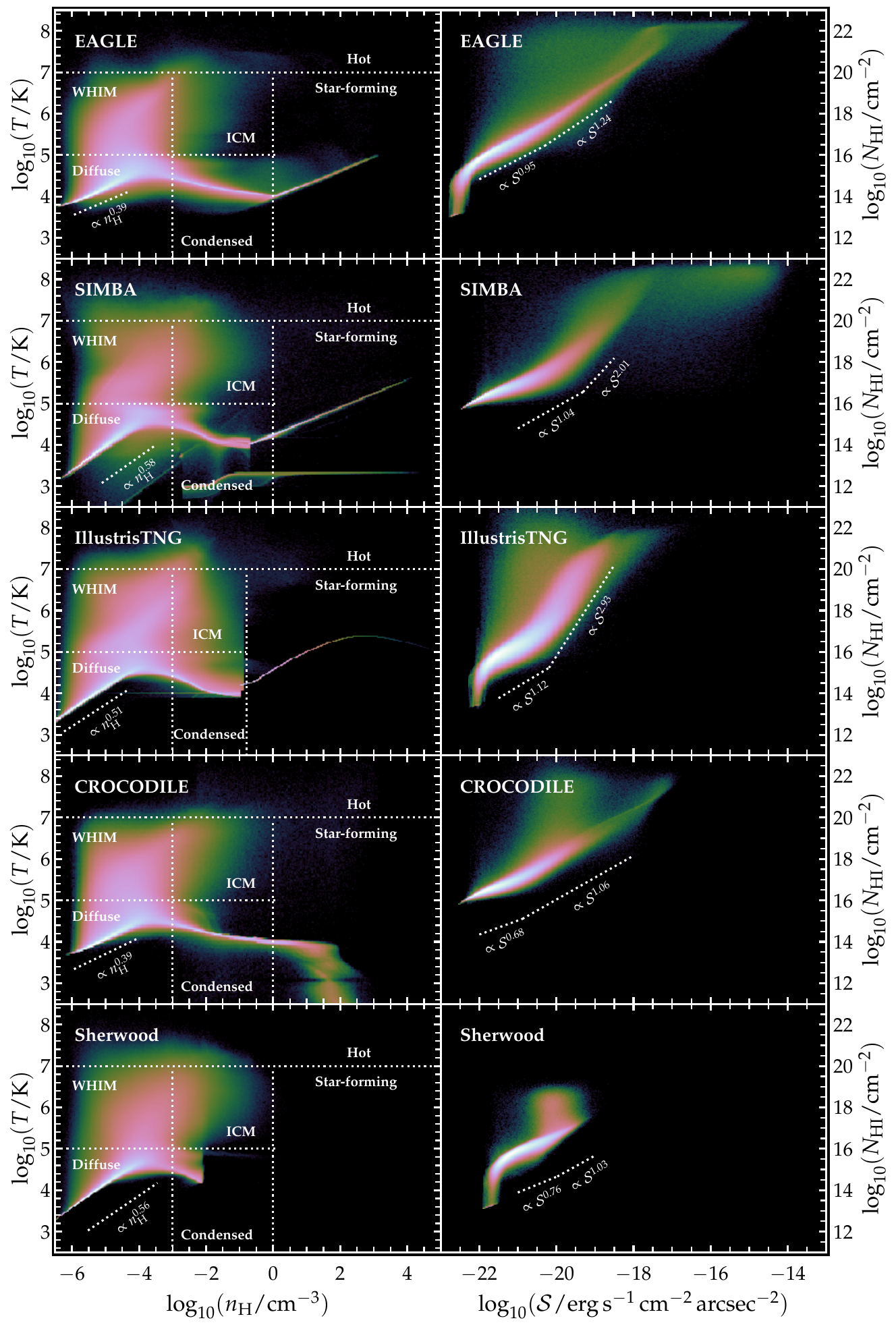}
    \caption{Left column (a): Temperature vs. hydrogen number density relation. Note the abrupt $n_{\rm H}$ cutoff for some simulations, a consequence of the switch to a stochastic Kennicutt–Schmidt star formation law. Right column (b): \HI{} column density vs. Lyman-$\alpha$ surface brightness. Power-law fits for this relation are shown as dotted lines, offset from the image data for visualisation. See Sections \ref{sec:T-lognH} to \ref{sec:insights}. The colour scale represents the number density of points contributing to each pixel in the plot. In the left column the range is approximately $0-10^6$ and for the right column it is approximately $0-10^4$.}
    \label{fig:nH-T_SB-NHI}
\end{figure*}

The right column of Fig. \ref{fig:nH-T_SB-NHI} illustrates the $\log_{10}\mathcal{S}$-$\log_{10} N_{\rm \HI{}}$ relations for each of the five simulations, which again reveal complex shapes. At low surface brightness, around the particle mass resolution, turn-downs are seen in EAGLE, IllustrisTNG and Sherwood. Above the particle resolution limit, the relationship approximates a broken power law behaviour, which are parameterised using
\begin{equation}
    N_{\rm \HI{}}(\rm \mathcal{S})=\begin{cases}
        N_0\left(\frac{\mathcal{S}}{\mathcal{S}_0}\right)^{\xi_1},\;\;\mathrm{for\;}\mathcal{S}<\mathcal{S}_0\\
        N_0\left(\frac{\mathcal{S}}{\mathcal{S}_0}\right)^{\xi_2},\;\;\mathrm{for\;}\mathcal{S}\geq\mathcal{S}_0,
    \end{cases}
    \label{eq:SB-NHI}
\end{equation}
where the constant $N_0$ is in units of $\mathrm{cm^{-2}}$. Fitting was carried out using the same approach described in Section \ref{sec:T-lognH}. Table \ref{tab:powerlawfits} illustrates the numerical results derived from each simulation, as well as the fitting ranges used.

\begin{table*}
\centering
\caption{Power law parameters (Eqs. \ref{eq:nh-T} and \ref{eq:SB-NHI}). See Fig. \ref{fig:nH-T_SB-NHI} and the discussion in Section \ref{sec:insights}. Here $\rm SB=erg\;s^{-1}\;arcsec^{-2}\;cm^{-2}$.}
\label{tab:powerlawfits}
\begin{tabular}{lrrlrrrrl}
\hline\hline
Simulation    & $\gamma$ & $T_0/10^{4}$ &  $n_{\rm H}/10^{-6} $ & $N_0/10^{17}$ & $\mathcal{S}_0/10^{-21}$ & $\xi_1$ & $\xi_2$ & $\mathcal{S}/10^{-21}$\\
& & [$\rm K$] & [$\rm cm^{-3}$] & [$\rm cm^{-2}$] & $[\rm SB]$ & & & $[\rm SB]$ \\ 
\hline
EAGLE         & $1.399 \pm 0.003$  & $1.462 \pm 0.008$& $1-100$ & $2.760\pm 0.183$ & $4.923\pm 0.289$ & $0.9525 \pm 0.005$ & $1.2426 \pm 0.004$ & $0.1-500$ \\
SIMBA         & $1.583 \pm 0.004$ &$0.315 \pm 0.002$ & $3-300$ & $36.014 \pm 6.674$  & $48.955 \pm 5.835$ & $1.0435 \pm 0.035$ & $2.0141 \pm 0.049$ & $1-1000$ \\
IllustrisTNG & $1.480 \pm 0.005$ & $0.979 \pm 0.010$ & $0.5-50$& $1.709\pm 0.267$ & $6.421\pm 0.487$ & $1.1218 \pm 0.036$ & $2.9317 \pm 0.066$ & $1-100$ \\
CROCODILE     & $1.396 \pm 0.004$ & $1.198 \pm 0.007$ & $1-50$& $1.276\pm 0.854$ & $1.316\pm 0.094$ & $0.6765 \pm  0.015$ & $1.0598 \pm 0.006$ & $1-100$ \\
Sherwood    & $1.516 \pm 0.003$ & $1.210 \pm 0.007$ & $2-100$& $0.4610 \pm 2.398$ & $10.589 \pm 0.641$ & $0.757 \pm 0.006$ & $1.034 \pm 0.015$ & $0.2-50$ \\
\hline\hline
\end{tabular}
\end{table*}

For $\mathcal{S}<\mathcal{S}_0$, from Table \ref{tab:powerlawfits}, $\bar{\xi_1} = 0.89 \pm 0.004$, which can be compared with a simple analytic prediction based on Lyman-$\alpha$ forest emission. Since $\xi_1$ concerns $\mathcal{S}<\mathcal{S}_0$, it relates to clouds with column densities $\log_{10}\mathrm{N_{\rm \HI}} \lesssim 18$ i.e. including Lyman limit systems but not damped Lyman-$\alpha$ systems (as the right column of Fig. \ref{fig:nH-T_SB-NHI} illustrates). 

The scaling of the Ly$\alpha$ surface brightness $S$ with neutral hydrogen column density $N_{\rm \HI}$ for optically thin forest absorbers can be estimated analytically. For recombination-dominated emission,
\begin{equation}
S \propto \alpha_{\rm B}\, n_e n_p\, L \simeq \alpha_{\rm B}\, n_{\rm H}^2\, L ,
\end{equation}
where $n_{\rm H}$ is the total hydrogen number density. Photoionisation equilibrium implies $n_{\rm \HI} \propto \alpha n_{\rm H}^2 / \Gamma$, so that
\begin{equation}
n_{\rm \HI} \propto \frac{\alpha_{\rm B}\, n_{\rm H}^2}{\Gamma},
\end{equation}
\begin{equation}
N_{\rm \HI} = n_{\rm \HI} L \propto \frac{\alpha_{\rm B}}{\Gamma}\, n_{\rm H}^2\, L .
\end{equation}
If absorbers are in local hydrostatic equilibrium, their characteristic thickness is of order the Jeans length,
\begin{equation}
L \sim L_J \propto T^{1/2} n_{\rm H}^{-1/2},
\end{equation}
leading to
\begin{equation}
N_{\rm \HI} \propto \frac{\alpha_B}{\Gamma}\, T^{1/2}\, n_{\rm H}^{3/2},
\qquad
S \propto \alpha_{\rm B}\, T^{1/2}\, n_{\rm H}^{3/2},
\end{equation}
recovering the \cite{Schaye2001} scaling for Ly$\alpha$ forest absorbers. To leading order, this predicts a near-linear relation $S \propto N_{\rm \HI}$, modulo weak temperature dependences of the recombination coefficient and spatial variations in the photoionisation rate $\Gamma$.

The fitted low-$S$ slopes found here ($\xi_1 \simeq 0.89$) are therefore close to, but slightly flatter than, the simplest optically thin hydrostatic expectation. Such a modest deviation is plausibly attributable to temperature-density variations, projection and multiphase mixing within the simulation volume, and departures from a single characteristic absorber scale. At higher column densities, where self-shielding, collisional excitation, and circumgalactic or galactic gas become important, this simple scaling is expected to break down and is likely to be the cause of the steeper behaviour discussed in the following subsection.

\subsection{Lyman-$\alpha$ surface brightness maps}\label{sec:SB}

\begin{figure*}
    \centering
    \includegraphics[width= \linewidth]{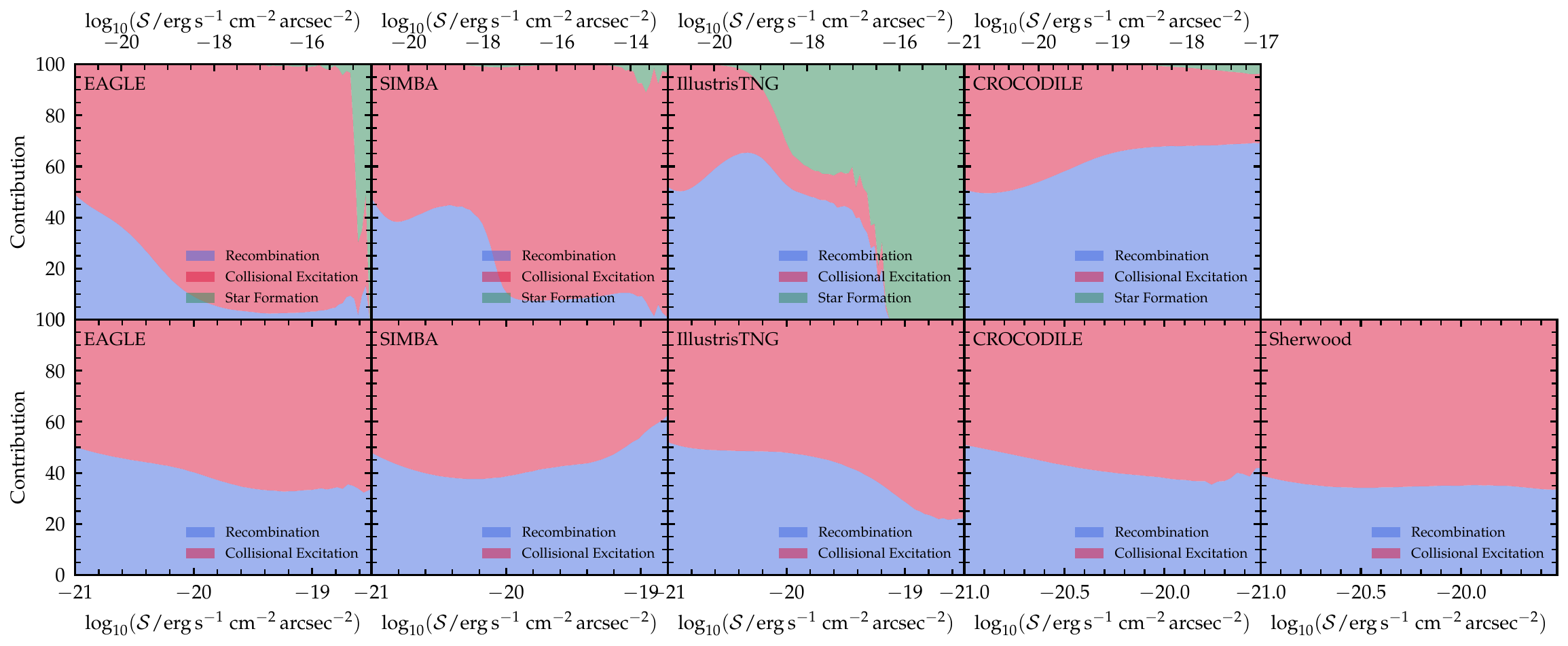}
    \caption{This illustrates the relative contributions towards the total surface brightness (for the models without adding background noise). The top row corresponds to the dust model of Section \ref{sec:dust}. The panel for Sherwood is missing for the reasons explained at the end of Section \ref{sec:addnoise}. The lower panel is for the model based on the number density limit (Section \ref{sec:hireject}).}
    \label{fig:contributions}
\end{figure*}

\begin{figure*}
    \centering
    \includegraphics[width=0.75\linewidth]{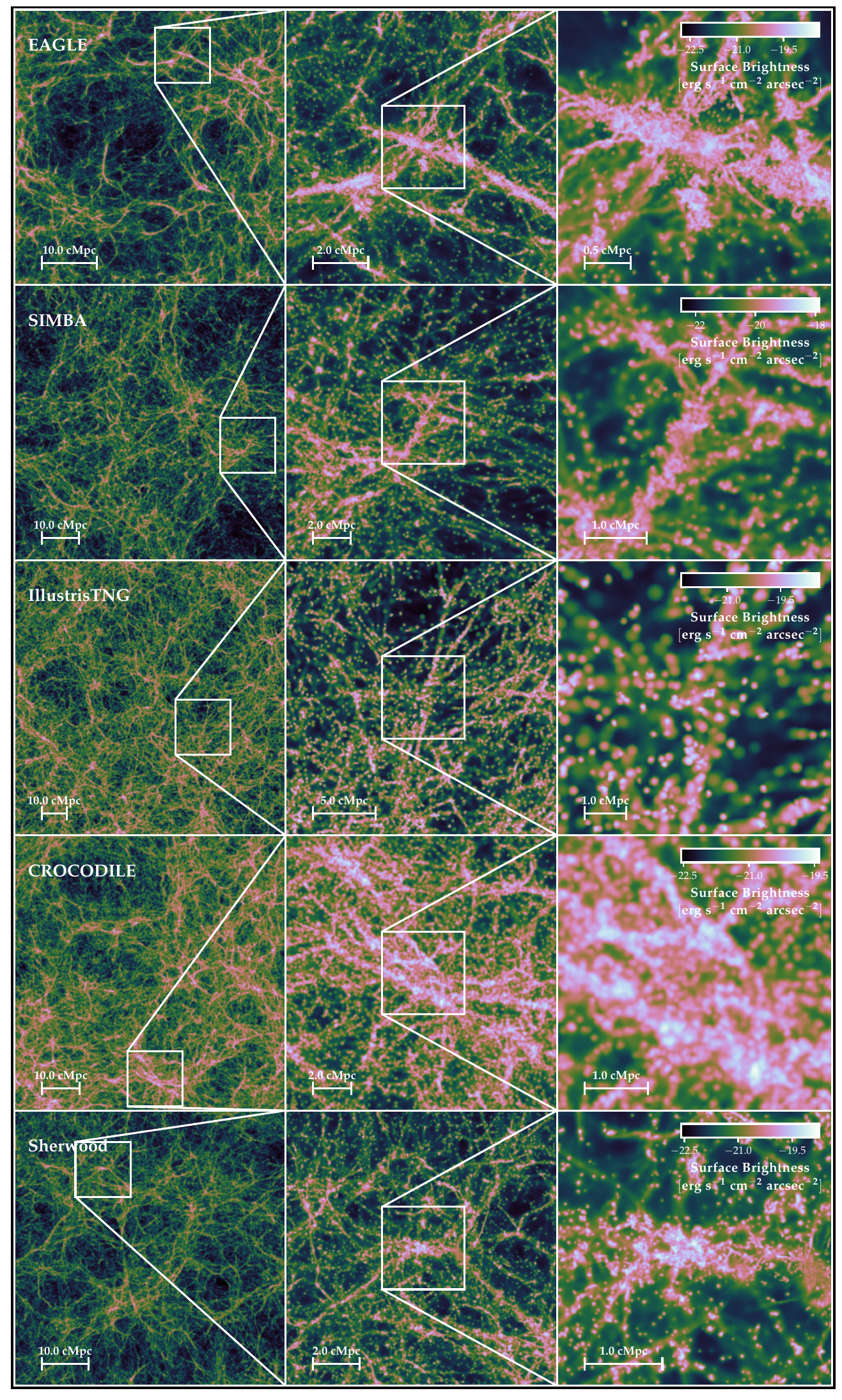}
    \caption{Lyman-$\alpha$ surface brightness map for each simulation with a particle density limit (see  Section \ref{sec:hireject}). Middle and right panels show 5$\times$ and 10$\times$ zoom-ins on arbitrarily selected regions. See Section \ref{sec:SB}. Each simulation has its own colour map, i.e. the numerical display ranges are not the same for each simulation. Instead, each colour map is set by the minimum and maximum count in each image.
}
    \label{fig:no_noise_maps}
\end{figure*}

\begin{figure*}
    \centering
    \includegraphics[width=0.75\linewidth]{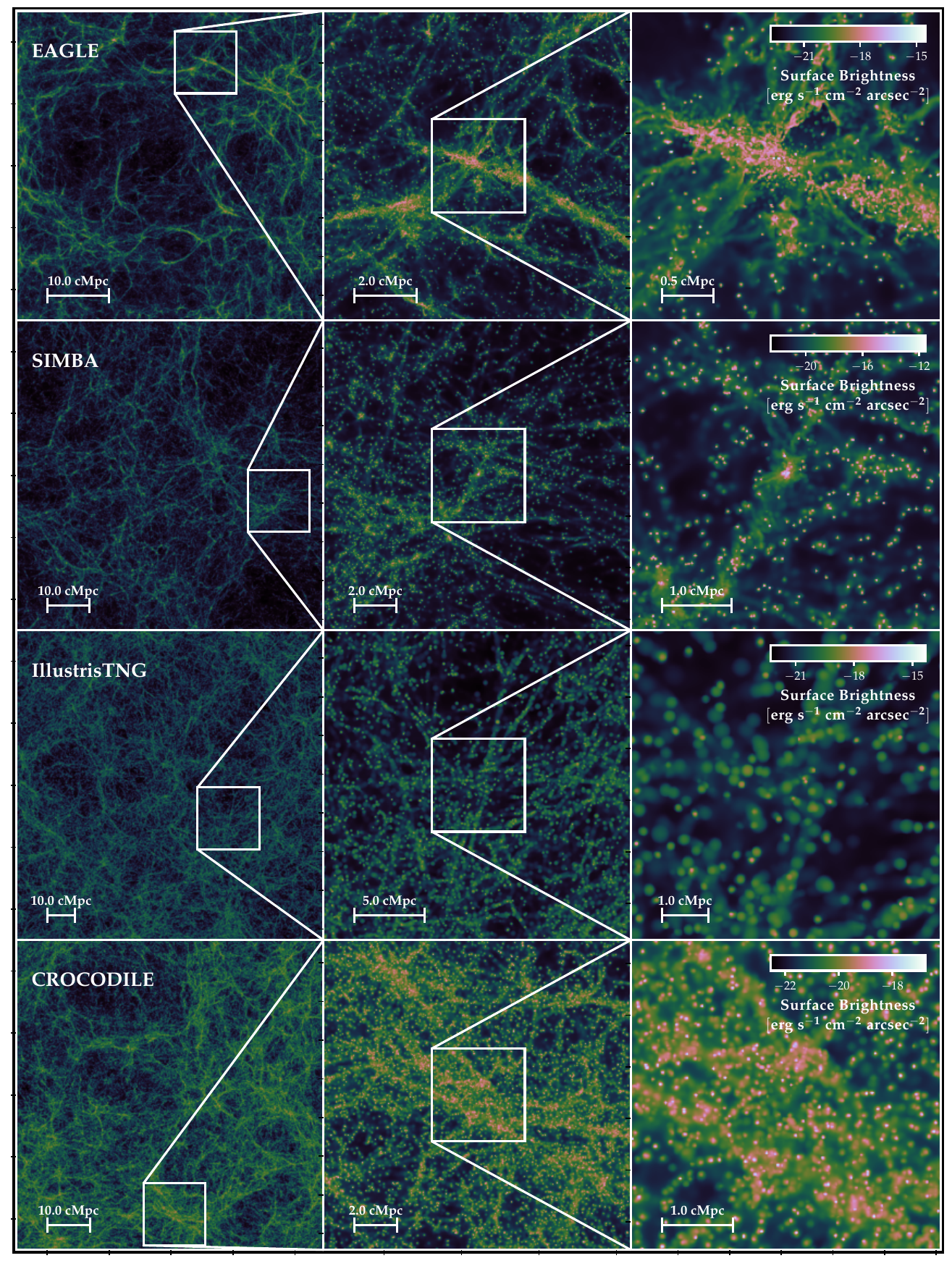}
    \caption{As Fig. \ref{fig:no_noise_maps} except here we use the dust model of Section \ref{sec:dust}. Sherwood is necessarily excluded for the reasons explained in Section \ref{sec:addnoise}.
}
    \label{fig:no_noise_maps_dust}
\end{figure*}

Fig. \ref{fig:no_noise_maps} presents the Lyman-$\alpha$ surface brightness maps for each of the five simulations, computed using the particle threshold prescription described in Section \ref{sec:hireject}. In Fig. \ref{fig:no_noise_maps_dust}, the corresponding maps are illustrated for the dust model of Section \ref{sec:dust}. The image size (in cMpc) depends on the simulation and is given in Table \ref{tab:5sims}, and converts to an angular scale on the sky using the cosmological parameters and redshift also given in Table \ref{tab:5sims}. The pixel size in each simulation image is closely matched to the Condor NM data described in Section \ref{sec:Intro}, with the overall image size being $4096\times 4096$ pixels. The middle and right columns give 5$\times$ and 10$\times$ zoom-ins on an arbitrary portion of each image. No background noise is added to these images. The intensity scaling is kept the same for all panels, such that the quite different results obtained from each simulation are visually apparent. The relative Lyman-$\alpha$ surface brightness contributions, for collisional excitation, recombination, and star formation, are illustrated in Fig. \ref{fig:contributions} (for the no-background noise model). The summed surface brightness histograms of pixel intensities for these five images are shown in the upper left panel of Fig. \ref{fig:SB_PDFs}, and the corresponding probability distribution functions are plotted in Fig. \ref{fig:SB_PDFs_decomposed} (again for the no-background noise model). These maps are calculated using the SMC dust model, using metallicity for each particle as provided by the simulation output, and hence the escape fraction is calculated (Eq. \ref{eq:f_esc}) on a particle-by-particle basis. 

The right panels of Fig. \ref{fig:nH-T_SB-NHI} show that the bulk of Lyman-$\alpha$ photons originate from clouds with $15 \lesssim \log_{10}\rm N_{\rm \HI} \lesssim 18$, concentrating the surface brightness signal in the range $-22 \lesssim \log_{10} \mathcal{S} \lesssim -19$, the faint end cut-off being determined by the simulation resolution. The column density along any sight-line is dominated by the {\it integrated} contributions from lower column density clouds. This surface brightness range is currently beyond detection. Therefore, for the purposes of deriving the Lyman-$\alpha$ surface brightness distribution, it is acceptable to use a simplified model of star/galaxy formation, i.e. all five simulations considered in this paper provide meaningful comparisons with the observational data; see the related discussions in Sections \ref{sec:T-lognH}-\ref{sec:insights}. 

In calculating the surface brightness distributions, we simply projected Lyman-$\alpha$ luminosities onto a 2D plane without using radiative transfer. Resonant Lyman-$\alpha$ scattering can, however, impact filament morphology and hence change surface brightness properties. The following approximation suggests that the effect could be substantial, in lower density regions particularly, but only for gas clouds where the line centre is saturated i.e. its optical depth $\tau_0 \ge 1$, and $\log_{10} N_{\rm \HI} \gtrsim 14$, which is applicable in our calculations (see Fig. \ref{fig:nH-T_SB-NHI}). We can combine Equations\,\eqref{eq:Jeans} and \eqref{eq:NHI} in $n_{\rm \HI} \approx N_{\rm \HI}/L$. Assuming a random walk, the transverse broadening of a filament can be represented by
\begin{equation}
    \Delta R \sim N_{\rm sc}^{1/2} l_0
\end{equation}
where $l_0 = 1/(n_{\rm \HI}\sigma_0)$ is the mean free path at the Lyman-$\alpha$ line centre, $\sigma_0$ is the absorption cross-section at the line centre ($\approx 5 \times 10^{-14}$ cm$^2$), and $N_{\rm sc} \sim \tau_0$ is the number of scatterings per photon, where $\tau_0 \approx N_{\rm \HI}\sigma_0$. We then have
\begin{equation}
    \Delta R \approx \frac{N_{\rm \HI}^{1/2}}{n_{\rm \HI}\sigma_0^{1/2}}
\end{equation}
We can get an idea of the impact of scattering (ignoring dust), adopting approximate quantities $T = 10^{4}\ \mathrm{K}$, $\Gamma = 10^{-12}\ \mathrm{s^{-1}}$, $\sigma_{0} = 5\times10^{-14}\ \mathrm{cm^{2}}$. Then, 
if $n_{\mathrm{H}} \sim 10^{-5}\ \mathrm{cm^{-3}}$ (very diffuse IGM), $\Delta R \sim 150\ \mathrm{kpc}$; 
if $n_{\mathrm{H}} \sim 3\times10^{-5}\ \mathrm{cm^{-3}}$, $\Delta R \sim 40\ \mathrm{kpc}$;
if $n_{\mathrm{H}} \sim 10^{-4}\ \mathrm{cm^{-3}}$ (denser filament/near-halo gas), $\Delta R \sim 9\ \mathrm{kpc}$;
if $n_{\mathrm{H}} \sim 3\times10^{-4}\ \mathrm{cm^{-3}}$ (yet denser clump), $\Delta R \sim 2\ \mathrm{kpc}$. 
Convert these to angular sizes. At $z \approx 2.48$, 1 arcsec $\approx 8.2$ kpc, so
$150 \mathrm{kpc} \rightarrow 18$'', 
$40 \mathrm{kpc} \rightarrow 5$'', 
$9 \mathrm{kpc} \rightarrow 1.1$'', 
$2 \mathrm{kpc} \rightarrow0.25$''.
Typical broadening could therefore reach a few kpc up to a few tens of kpc, i.e. $\sim 0.3''$ up to a few tens of arcseconds, depending strongly on the filament density and column density. These considerations, although crude, may have interesting implications for future work, particularly when high-quality cosmic web images have been obtained, and statistics such as the image autocorrelation function might provide stringent cosmological and physical constraints.

\subsubsection{Adding background noise - emulating real observations}\label{sec:addnoise}

\begin{figure*}
    \centering
    \includegraphics[width=0.85\linewidth]{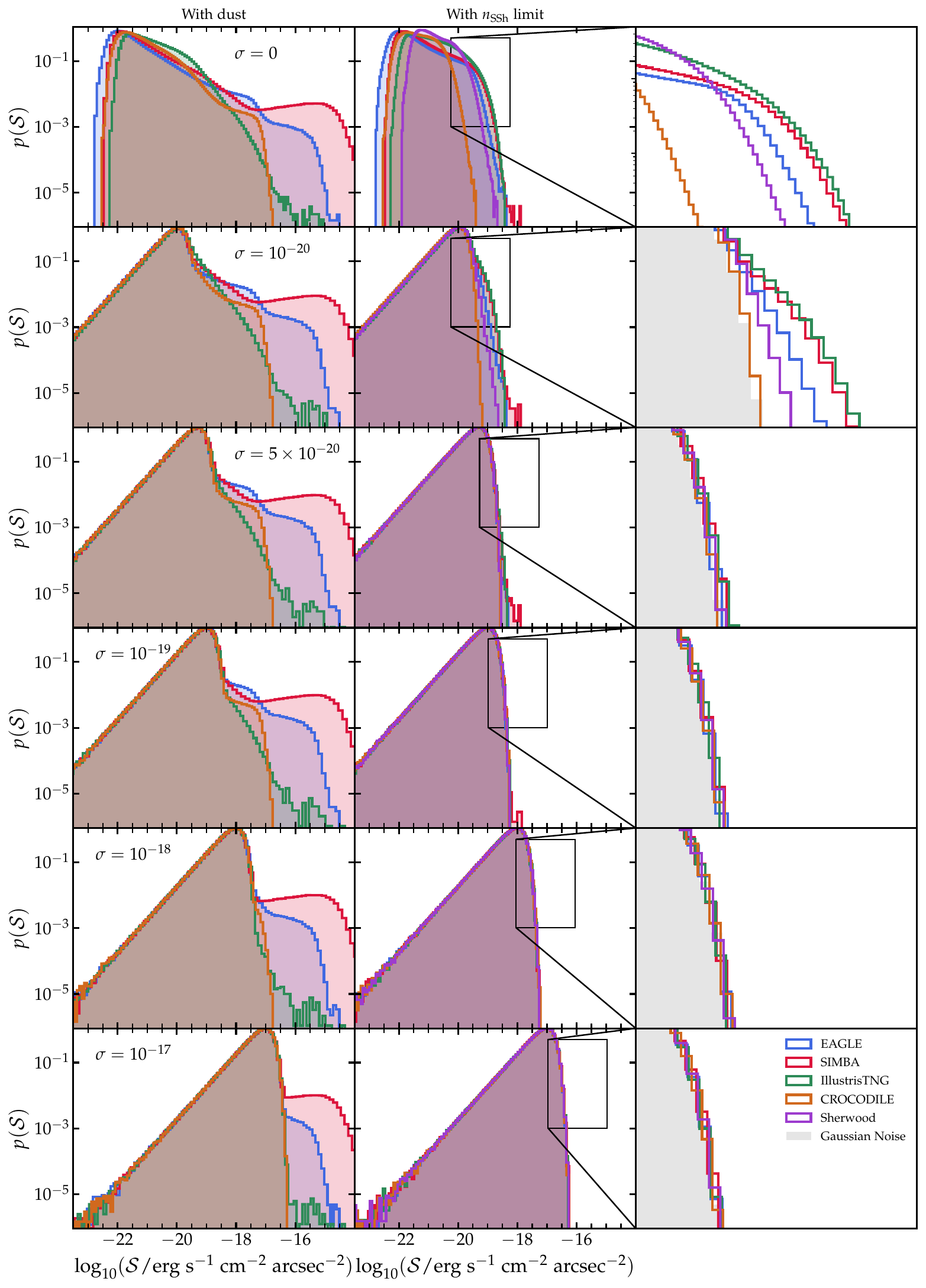}
    \caption{Lyman-$\alpha$ narrow-band surface brightness in each simulation. Each panel shows the surface brightness probability distribution. The left column corresponds to the dust models described in Section \ref{sec:dust}. The standard deviation $\sigma$ of the Gaussian noise added is shown in each panel. The top row ($\sigma = 0$) has no noise. The middle and right panels correspond to the particle threshold method described in Section \ref{sec:hireject}. The left column (from which Sherwood is excluded, since metallicity information is not available) illustrates that the combined intergalactic and circumgalactic narrow band Lyman-$\alpha$ is detectable by all simulations for $\mathcal{S} \sim 10^{-18}$ erg s$^{-1}$ cm$^{-2}$ arcsec$^{-2}$. The middle column illustrates that detecting the {\it low density} component of the cosmic web narrow band Lyman-$\alpha$ requires a background noise level below $\sigma \sim 5 \times 10^{-20}$ erg s$^{-1}$ cm$^{-2}$ arcsec$^{-2}$. Fig. \ref{fig:Asquared} shows the Anderson-Darling statistical test applied to these data, to evaluate detection thresholds.}
    \label{fig:SB_PDFs}
\end{figure*}

\begin{figure*}
    \centering
    \includegraphics[width=\linewidth]{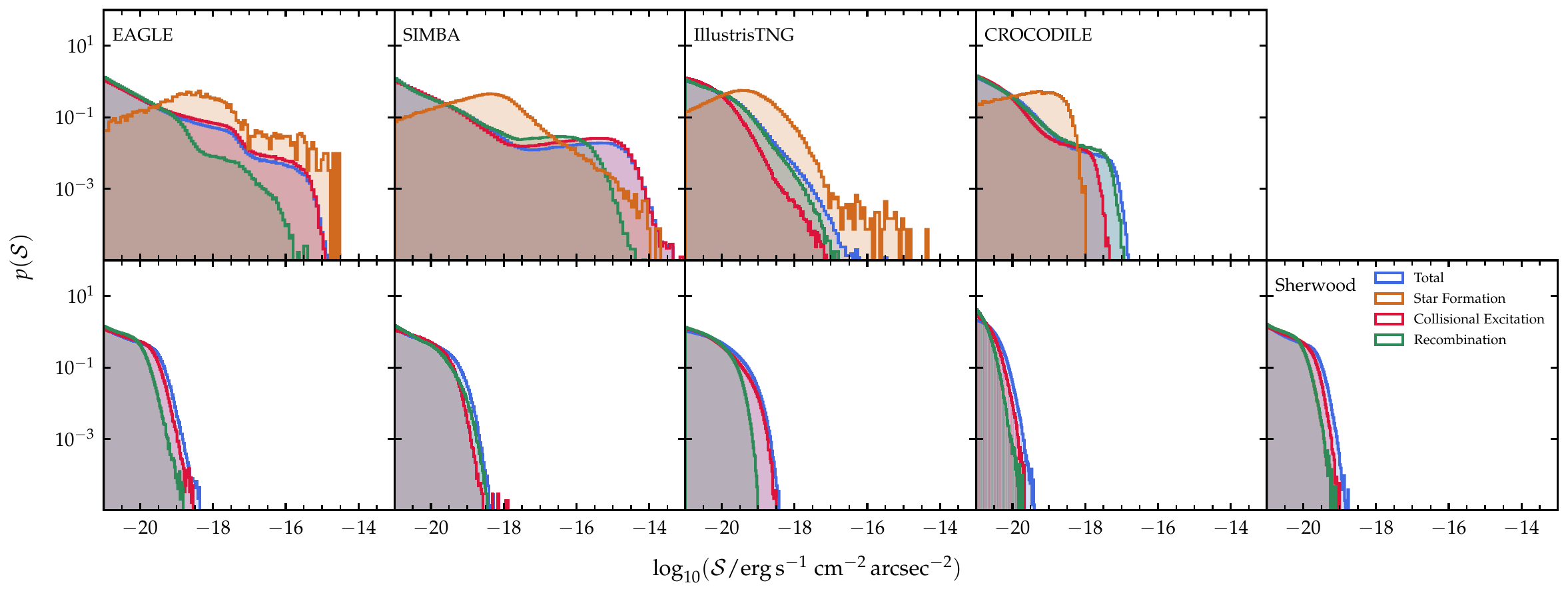}
    \caption{Surface brightness probability distribution functions for each simulation, illustrating the relative contributions from collisional excitation, recombination, and star formation processes. This figure corresponds to models without adding background noise, and thus it may be compared with Fig. \ref{fig:contributions}.}
    \label{fig:SB_PDFs_decomposed}
\end{figure*}

\begin{figure*}
    \centering
    \includegraphics[width=1.02\linewidth]{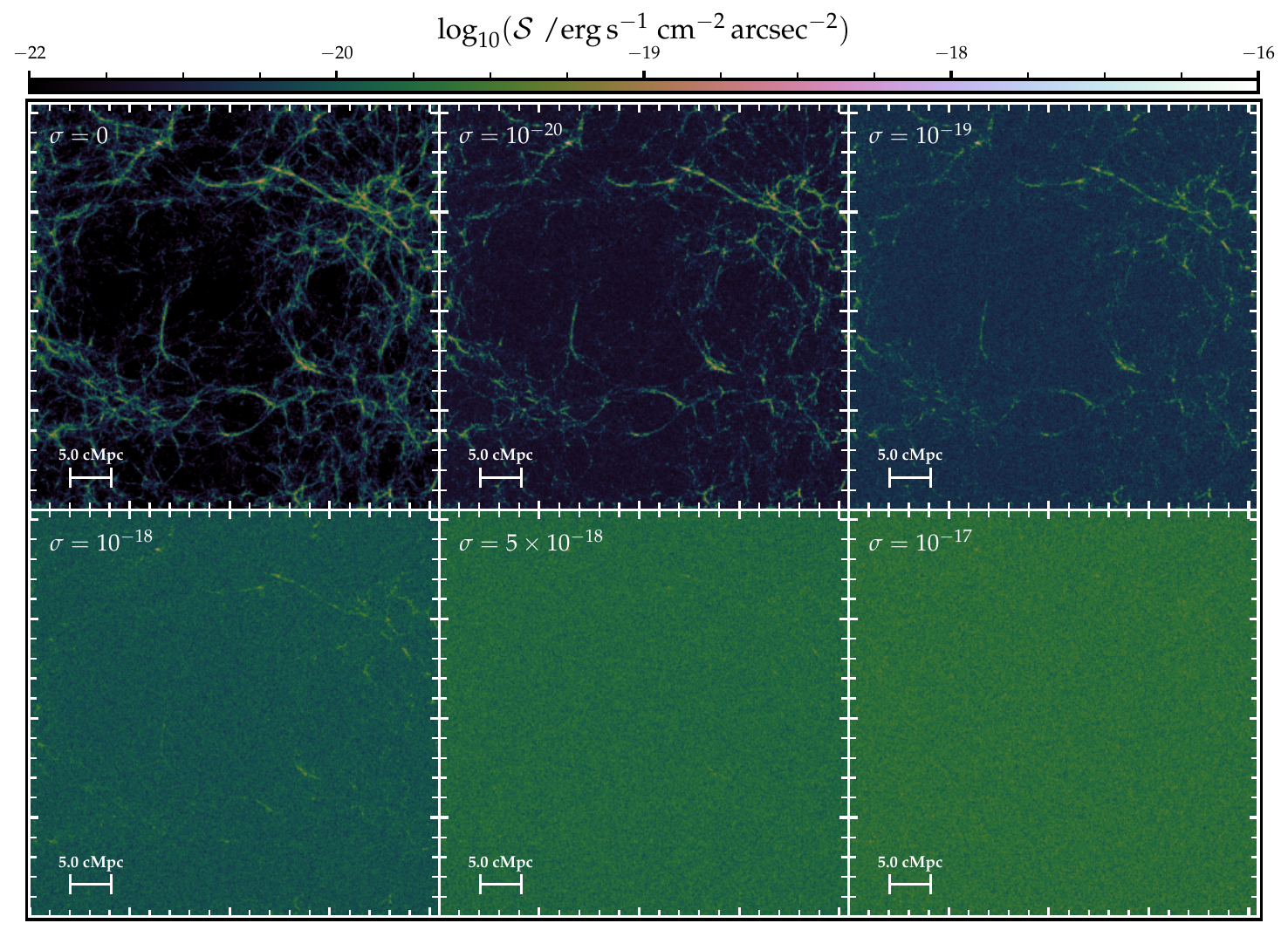}
    \caption{Surface brightness map from the EAGLE simulation made using the dust calculation (Section \ref{sec:dust}), for various Gaussian noise models (Section \ref{sec:addnoise}), emulating real observational data. The standard deviation $\sigma$ of the Gaussian noise added is shown in each panel. The panels in this figure correspond to the left-hand panels in Fig. \ref{fig:SB_PDFs}. 
    }
    \label{fig:noisy_maps}
\end{figure*}

Figs. \ref{fig:SB_PDFs} and \ref{fig:noisy_maps} illustrate the results from adding Gaussian noise of varying $\sigma$, representing a broad range of potential observational measurements. Narrow-band Lyman-$\alpha$ emission from the diffuse cosmic web gas, Lyman-$\alpha$ emitters, galaxies, quasars, etc., will show up as a UV excess in the high $\mathcal{S}$ tail of the intensity distribution function for the combined noise+simulation data. Fig. \ref{fig:SB_PDFs} shows these distribution functions for each cosmological simulation. 

In the left column of Fig. \ref{fig:SB_PDFs} (where the dust model of Eq. \ref{eq:f_esc} has been applied without culling high optical depth simulation particles), we see diverse results. It is clear that such detections would be associated with emission from higher optical depth particles due to the lack of detection in the middle column of Fig. \ref{fig:SB_PDFs}. Other interesting features can be seen in Fig. \ref{fig:SB_PDFs}. Considering the top row only, middle panel (particle threshold case), very diverse fall-offs at low $\log_{10} \mathcal{S}$ are seen. This effect is presumably associated with differing simulation resolutions, differing gas physics treatments, different redshifts for each simulation, and other factors. More consistent fall-offs are seen at the high flux end (for the top row, where no noise has been added), with the exception of CROCODILE. We avoid speculating further on this point.

Fig. \ref{fig:noisy_maps} illustrates the EAGLE surface brightness map for various realisations of the background noise, from $\sigma = 10^{-20}$ up to $10^{-17} \;\rm erg \; s^{-1} \; cm^{-2} \; arcsec^{-2}$. The top left panel shows the noise-free map. An eye inspection of these panels indicates that a background noise of $\sigma \sim 10^{-20}$ or better is required to reveal a conspicuous cosmic web pattern. 

To test for flux deviations in the tail of the dominant noise contribution, we apply the Anderson-Darling statistic \citep{Anderson1952}. We first add Gaussian noise to the final surface brightness map, Eq. \eqref{eq:S_final}, and form its cumulative distribution function, $P(\mathcal{S})$. The cumulative distribution function for the Gaussian noise alone is $F(\mathcal{S})$. The non-cumulative PDFs from which these cumulative distribution functions are formed are illustrated in Fig. \ref{fig:SB_PDFs}.

The Anderson-Darling statistic measures the weighted squared difference between the cumulative empirical $P(\mathcal{S})$ and model $F(\mathcal{S})$ distribution functions using
\begin{align}
A^2 & = n \int^\infty_{-\infty} (P(\mathcal{S})-F(\mathcal{S}))^2 \, w(\mathcal{S}) \, dF(\mathcal{S}) \\
    & = n \int^\infty_{-\infty} a_w(\mathcal{S}) \, dF(\mathcal{S}),
\label{eq:Asquared}
\end{align}
where $n$ is the number of points in the sample. The weights
\begin{equation}
w(\mathcal{S}) = 1/[F(\mathcal{S})\,(1-F(\mathcal{S}))]
\end{equation}
serve the purpose of applying extra weight in the tails, since $w(\mathcal{S})$ becomes large when $F(\mathcal{S})$ is close to $0$ or $1$ (i.e., the extreme left or right tail). This is helpful for our specific application, as we are looking for a small surface brightness excess in the upper wing of a dominant noise background. Small discrepancies in the bulk of the distribution contribute only modestly, whereas even tiny departures in the tails are amplified in the sum over the distribution. 

\begin{figure*}
    \centering
    \includegraphics[width= 0.8\linewidth]{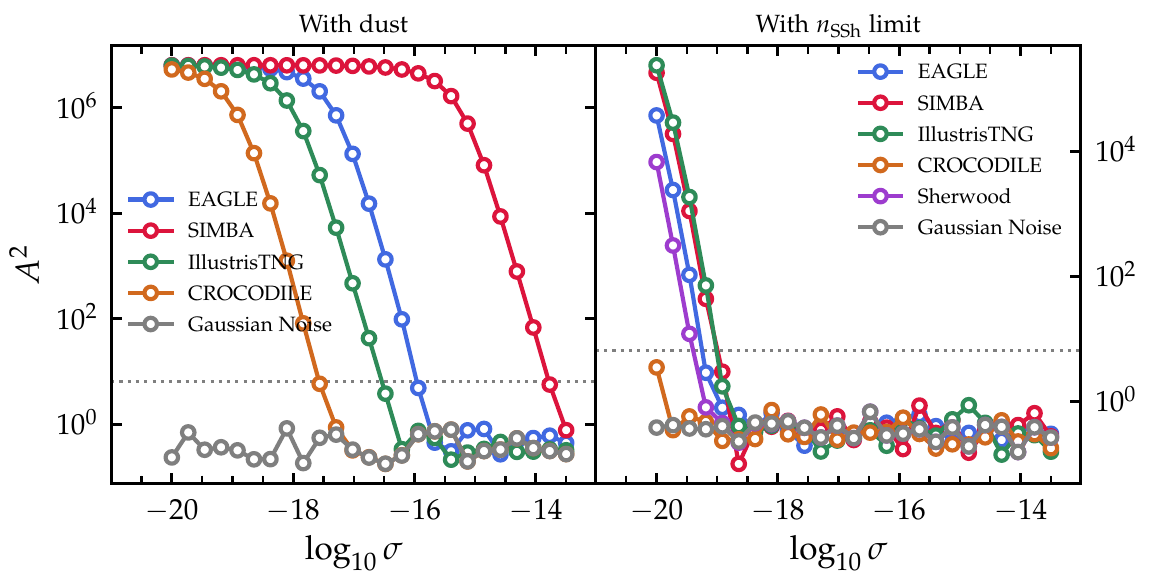}
    \caption{Anderson-Darling $A^2$ statistic (Eq. \ref{eq:Asquared}) for each simulation, for different values of added Gaussian noise $\sigma$. The left panel here corresponds to the left panel in Fig. \ref{fig:SB_PDFs}. The right panel corresponds to the central and right panels in Fig. \ref{fig:SB_PDFs}. The horizontal dotted line indicates a 5-standard-deviation detection threshold (provided by the \texttt{SciPy} package \protect\citep{Virtanen2020}). This statistic therefore tells us what level of noise, in real astronomical observations, must be reached for a 5-$\sigma$ detection of wide-field Lyman-$\alpha$ emission from the cosmic web, as expected according to each simulation depicted. See Section \ref{sec:addnoise}.
    }
    \label{fig:Asquared}
\end{figure*}

Fig. \ref{fig:Asquared} shows $A^2$ as a function of Gaussian background noise for each simulation. The left panel shows the result of a calculation where {\it no} high-density particle removal is carried out, and instead the dust calculation described in Section \ref{sec:dust} is used. The right panel illustrates the results when no dust calculation is applied, and instead high-density particles are removed, as described in Section \ref{sec:hireject}. Unsurprisingly, the left and right panels show dramatically different results; the right panel illustrates the detectability of low-density Lyman forest emission, whilst the left panel shows the detectability of all Lyman-$\alpha$ emission, allowing for dust attenuation. The right panel shows that the most optimistic detection of the low-density cosmic web (IllustrisTNG) requires a sensitivity of $\sim 2 \times 10^{-19}$ erg s$^{-1}$ cm$^{-2}$ arcsec$^{-2}$. Sherwood is not shown in the left column of Fig. \ref{fig:Asquared} because the version used does not provide particle metallicities, so the dust calculation cannot be done. The remaining four simulations agree that excess narrow-band Lyman-$\alpha$ emission should be detected above the 5$\sigma$ level for fluxes brighter than $\sim 8 \times 10^{-17}$ erg s$^{-1}$ cm$^{-2}$ arcsec$^{-2}$.

It is interesting to compare the results just discussed with the particularly bright Lyman-$\alpha$ emitter reported in \cite{Cantalupo2014}, for which the peak Lyman-$\alpha$ surface brightness is $\sim 10^{-16}$ erg s$^{-1}$ cm$^{-2}$ arcsec$^{-2}$. Fig. \ref{fig:Asquared} evidently shows a very large range for the four simulations considered in the dust model, $ -17.5 < \log_{10} < -14$, although it is noteworthy that EAGLE and IllustrisTNG fall rather close to this observational result.

\section{Summary and Outlook}\label{sec:Conclusions}

In this work, we use post-processing of five cosmological simulations to predict narrow-band Lyman-$\alpha$ emission from the cosmic web at redshift $z \approx 2.5$, incorporating the effects of hydrodynamic processes and dust attenuation. The consistency checks obtained by cross-comparing simulation results suggest our results are robust. Quantitative estimates are given for the faint extended surface brightness that future observations will be able to detect for achievable background noise levels. We summarise as follows:
\begin{enumerate}
  \item We post-process five hydrodynamic cosmological simulations, computed at different redshifts in the range $2.00<z<2.74$ and with different comoving box sizes (Table \ref{tab:5sims}). The data are used to calculate five cosmic web Lyman-$\alpha$ surface brightness maps. 
  \item Each simulation invokes physics differently, so the surface brightness results are cross-compared to evaluate consistency.
  \item From each simulation, we also derive the {\HI} column density distribution $f(N_{\rm  \HI}, X)$, and compare the results against a compilation of quasar absorption measurements, covering the approximate neutral hydrogen column density ranges $14.5 < \log_{10} N_{\rm  \HI} < 17.5$ and $19 < \log_{10} N_{\rm  \HI} < 22.3$, with an observational gap in the middle corresponding to the Lyman limit to sub-DLA range. All five simulations give good agreement for $\log_{10} N_{\rm  \HI} < 17.5$. Three simulations (EAGLE, SIMBA, IllustrisTNG) also agree fairly well at $\log_{10} N_{\rm  \HI} > 19$, but CROCODILE underestimates $f(N_{\rm \HI}, X)$ and Sherwood is designed to model the low density cosmic web component so is not applicable.
  \item We explore two approaches for handling high-density simulation particles, where photon scattering and dust effects are expected to attenuate the Lyman-$\alpha$ escape fraction. The simplest approach is to discard all simulation particles exceeding a density threshold, based on a Jeans-limit gas cloud size. The second approach retains all simulation particles irrespective of density and calculates photon scattering and dust effects using an SMC dust model.
  \item From the data used to form the final surface brightness maps, we explore the $T-n_{\rm{H}}$ and $\mathcal{S}-N_{\rm \HI}$ relations. The low-density regions in the $T-n_{\rm{H}}$ plane show that the gas temperature predicted by the different simulations ranges by a factor of $\sim 6$. This should be detectable using Voigt profile modelling of Lyman forest absorption lines along sight lines through each simulation box.  The $\mathcal{S}-N_{ \rm  \HI}$ plane is similar to the expected scaling of the surface brightness with $\rm \HI{}$ column density for optically thin gas.
  \item We contrast procedural differences between real and simulated image processing. These differences become relevant when interpreting surface brightness measurements from both.
  \item We emulate (at least partially) real data by adding varying levels of Gaussian noise to the simulated Lyman-$\alpha$ surface brightness maps to evaluate the limiting background noise required of observations that may detect narrow-band emission.
  \item Deep, narrow-band, wide-field images targeting Lyman-$\alpha$ emission at some selected redshift are also exposed to wide-field emission lines at other redshifts. We offer simple arguments that suggest the strongest contaminant, [\OII], should be negligible (Appendix\,\ref{sec:OIInegligible}).
  \item The Anderson-Darling statistic provides an indication of the detectability of wide-field Lyman-$\alpha$ emission from the cosmic web, as predicted by the four simulations studied. For the dust models, SIMBA finds a substantially more optimistic prediction than EAGLE, IllustrisTNG, and CROCODILE. The wide dynamic range from the dust calculations seen in the Anderson-Darling statistic reflects the correspondingly diverse approaches taken across the simulations, at least for higher particle densities.
\end{enumerate}

Whilst hydrodynamic simulations have succeeded in reproducing some observational quantities (e.g. the $N_{\rm \HI}$ CDDF), the dust curves in Fig. \ref{fig:SB_PDFs} and $A^2$ in Fig. \ref{fig:Asquared} reveal huge variations, possibly indicating discrepant metallicity values across the simulations. Whilst the particle threshold approach (Section \ref{sec:hireject}) offers simplicity, the dust method has the in-principle appeal of predicting emission from beyond pure IGM Lyman-$\alpha$. The present work, however, exposes large prediction variations amongst the existing simulations (e.g. Figs. \ref{fig:SB_PDFs} and \ref{fig:Asquared}). 

Although different dust prescriptions can change surface-brightness predictions locally (especially in the circumgalactic medium and the bright end), the faint, diffuse emission that traces the cosmic web is much less sensitive to these details. Other studies likewise show that dust mainly redistributes light around galaxies without strongly affecting the faint-end statistics \citep{Vogelsberger2020}. In our work, we explored two contrasting approaches to high-density gas: applying an SMC-type dust prescription to all particles, and, at the other extreme, excluding optically thick particles. We compared results across five independent simulations. 

Our finding carries two major implications. First, we improve our predictive framework for assessing the detectability of the cosmic web in Lyman-$\alpha$ with current and planned wide-field facilities. Second, we provide strong theoretical backing for the emerging observational frontier represented by Condor and its future extensions in Chile, which will be capable of producing the first cartographic mapping of the cosmic web across cosmic time.

Looking forward, the combination of deep narrowband imaging, spectroscopic follow-up, and further refined simulations will enable us to disentangle the relative contributions of dust, gas dynamics, and radiative transfer to the observed morphology of filaments. Such advances will ultimately allow us not only to confirm the detection of the cosmic web in emission, but also to use its properties as a new probe of galaxy formation, large-scale structure in the Universe, and to help establish the precise cosmological model of our Universe.

\section*{Acknowledgments}
We are grateful to several people who kindly provided their expertise: Deryck Thake (for setting up IoA accounts and other computing details), Yuri Oku (for providing CROCODILE data), Romeel Dav\'e (for communications about SIMBA), Peter Laursen (for his help with MoCaLaTA), and Ewald Puchwein (for useful advice on Sherwood).

OS is grateful for funding provided by COST (European Cooperation in Science and Technology) Action CA21136 ``Addressing observational tensions in cosmology with systematics and fundamental physics (CosmoVerse)'', enabling an initial short-term visit to the IoA/Kavli, Cambridge University, where this work commenced, and also thanks the IoA for hosting an extended second visit in July-September 2025. 
JKW is grateful for a visiting position at ESO Santiago, May-July 2025, where part of this work was carried out. 
JW gratefully acknowledges support from the Cosmic Dawn Center through the DAWN Fellowship. The Cosmic Dawn Center (DAWN) is funded by the Danish National Research Foundation under grant No. 140.
GG thanks Pontificia Universidad Católica de Chile, ESO, France-Chile Laboratory of Astronomy (FCLA), and Laboratoire d’Astrophysique de Marseille (LAM) for their support during a 2024-2025 sabbatical leave.

The numerically intensive calculations were performed on the OzSTAR national facility at Swinburne University of Technology. OzSTAR is partly funded by the Astronomy National Collaborative Research Infrastructure Strategy (NCRIS) allocation provided by the Australian Government, and from the Victorian Higher Education State Investment Fund (VHESIF) provided by the Victorian Government. We are grateful for access to the computing facilities at the Institute of Astronomy, Cambridge, and for the availability of the open-source \texttt{Python} packages: \texttt{SciPy} \citep{Virtanen2020}, \texttt{Matplotlib} \citep{matplotlib}, and \texttt{NumPy} \citep{numpy}.

\section*{Data Availability}

This work is based solely on numerical simulations and does not use any observational data. The simulation data underlying this article will be shared upon reasonable request to the corresponding author.

\bibliographystyle{mnras}
\bibliography{simulations}

\appendix 

\section{Expected relative strengths of cosmic web emission from Lyman-$\alpha$ at $z=2.4754$ and [\OII] at $z=0.1332$}\label{sec:OIInegligible}

We can estimate the relative brightnesses of Lyman-$\alpha\;(z=2.4754)$ and [\OII] ($z=0.1332$) by combining three contributing factors as follows:\\
(i) {\it Relative abundances:}\\
The oxygen relative abundance is
\begin{equation}
\frac{n_{\rm O}}{n_{\rm H}} \sim \left(\frac{\rm O}{\rm H}\right)_\odot \left(\frac{Z}{Z_\odot}\right) \sim 4.9 \times 10^{-4} \left(\frac{Z}{Z_\odot}\right).
\end{equation}
Empirical constraints on the metallicity of the low-redshift cosmic web range over 3 orders of magnitude, from $10^{-3} \lesssim Z_\odot \lesssim 1$, depending on neutral hydrogen column density \citep{Shull2014, Werk2013}. For an illustrative calculation, we use a low redshift cosmic web metallicity $Z \sim 0.1\,Z_\odot$,
\begin{equation}
\frac{n_{\rm O}}{n_{\rm H}} \sim 5 \times 10^{-5},
\end{equation}
An over-estimate since we take the limiting case of all oxygen being singly ionised.\\
(ii) {\it Tolman surface brightness dimming:}\\
The observed surface brightness of a line emitted by diffuse gas at redshift $z$ scales as
$\mathcal{S}_{\rm obs} \propto \mathcal{S}_{\rm em}/(1+z)^4$ (see Eq. \eqref{eq:S_final} for the derivation of the surface brightness), where $\mathcal{S}_{\rm em}$ is the emitted surface brightness. For our narrow-band filter, Lyman-$\alpha$ is thus dimmed by a factor of 90 relative to [\OII], purely due to cosmological surface brightness dimming.\\
(iii) {\it Relative emissivities:}\\
Lyman-$\alpha$ emission from the diffuse cosmic web arises primarily from recombination and collisional excitation of neutral hydrogen and primarily where the gas density is lower ($n_{\rm H} \sim 10^{-4.5} - 10^{-3} \rm{cm}^{-3}$, more typical of the bulk of the diffuse cosmic web). [\OII] 3727 emission comes primarily from collisions, but this is efficient only in higher gas densities ($n_{\rm H} \sim 10^{-3} - 10^{-2}\,\rm{cm}^{-3}$) that do not form the bulk of the cosmic web. For Case~B hydrogen recombination (optically thick Lyman continuum, Lyman-$\alpha$ is resonantly scattered, effectively increasing the observed surface brightness along the line of sight), the emissivity is 
\begin{equation}
\epsilon_{\rm Ly\alpha} = h\nu_{\rm Ly\alpha}\,\alpha_{\rm eff}\,n_e n_p,
\label{eq:emisLya}
\end{equation} 
where $h\nu_{\rm Ly\alpha} = 10.2~{\rm eV} = 1.63 \times 10^{-11}~{\rm erg}$ and $\alpha_{\rm eff} \simeq 0.68\,\alpha_B \sim 1.77 \times 10^{-13}~{\rm cm^3\,s^{-1}}$ at $T = 10^4~{\rm K}$. Thus 
\begin{equation}
\epsilon_{\rm Ly\alpha} 
\approx 2.9 \times 10^{-24}\, n_e n_p \; {\rm erg\,cm^{-3}\,s^{-1}}.
\end{equation}
[\OII] is a forbidden, collisionally excited line. Putting $x_{\rm \OII} = n_{\rm \OII}/n_{\rm O}$, the volume emissivity is  
\begin{equation}
\epsilon_{\rm [\OII]} = n_{\rm \OII}\, n_e\, q_{ul}\, h\nu \approx x_{\rm \OII} \frac{n_{\rm O}}{n_{\rm H}}\, n_e n_p\, q_{ul}\, h\nu.
\label{eq:eps-OII}
\end{equation}
Using the same abundance over-estimate as above, we set $x_{\rm \OII}=1$ and use the approximation $n_p \sim n_e$, and use
\begin{equation}
q_{ul} = \frac{8.63 \times 10^{-6}}{T^{1/2}} \, \frac{\Omega_{ul}}{g_l}\, \exp\!\left(-\frac{E}{kT}\right)\quad \; [{\rm cm^3\,s^{-1}}].
\end{equation}
For [O\,II] $\lambda 3727$~\AA, $h\nu = 3.3~{\rm eV} = 5.3 \times 10^{-12}~{\rm erg}$, with collision strength $\Omega_{ul} \sim 1.5$, lower-level degeneracy $g_l = 4$, and energy gap $E \sim 3.3~{\rm eV}$. At $T = 10^4~{\rm K}$, $E/kT 
\sim 3.8$, so
$q_{ul} 
\sim 7 \times 10^{-10}~{\rm cm^3\,s^{-1}}$.
Putting those values into Eq. \eqref{eq:eps-OII}, the emissivity then becomes
\begin{equation}
\epsilon_{\rm [O\,II]} \sim 1.8 \times 10^{-25}\, n_e n_p\; {\rm erg\,cm^{-3}\,s^{-1}},
\end{equation}
and the emissivity ratio is thus
\begin{equation}
\frac{\epsilon_{\rm [\OII{}]}}{\epsilon_{\rm Ly\alpha}} \;\sim\; 0.062
\end{equation}

We note an approximation made above; the quantities $n_e$ and $n_p$ appearing above relate to the particle densities at $z=2.4754$ (Eq. \ref{eq:emisLya}) and $0.1332$ (Eq. \ref{eq:eps-OII}). We avoid scaling individual particle densities using simple Hubble dilutions of $(1+z)^3$, because that would ignore the competing effect of gravitational growth. Whilst there is considerable uncertainty in the latter, simulations suggest the two effects approximately cancel. For example, \cite{Bahe2025} find overdensities at $z=2$ and $z=0$ in the EAGLE and TNG100 simulations differ by a factor $\approx 2$. For this reason, we have made the approximation above that $n(z=2.4754) = n(z=0.1332)$. Finally, we may therefore combine the three factors above to estimate the relative strengths of [\rm \OII] ($z=0.1332$) and Lyman-$\alpha \;(z=2.4754)$ emission,
\begin{equation}
\frac{\mathcal{S}_{\rm [\OII]}}{\mathcal{S}_{\rm Ly\alpha}} \ll 5 \times 10^{-5} \times 90 \times 0.062 \sim 3 \times 10^{-4}.
\end{equation}
Therefore, we may expect that attempts to detect statistical wide-field Lyman-$\alpha$ cosmic web emission using narrow band imaging at $z \approx 2.5$ are likely to contain a negligible amount of contamination from low redshift ($z \approx 0.13$) [\rm \OII].\\
\vspace{0.2in}

\bsp
\label{lastpage}

\end{document}